\documentclass[onecolumn,superscriptaddress,floatfix,preprintnumbers,
nofootinbib,hyperref]{revtex4}
\pdfoutput=1
\usepackage[colorlinks=true,breaklinks=true]{hyperref}
\usepackage[normalem]{ulem}
\usepackage[utf8]{inputenc}
\hypersetup{allcolors=[rgb]{0.0 0.0 0.6},linkcolor=[rgb]{0.75 0.05
0.05}}
\usepackage{amsmath,amssymb}
\usepackage{epsfig}
\usepackage{graphicx}
\usepackage{slashed}
\usepackage{url}
\usepackage{color}
\usepackage{multirow}
\usepackage{amsfonts}
\usepackage{subfigure}
\usepackage{graphicx}
\usepackage{verbatim}
\usepackage[dvipsnames]{xcolor}
\usepackage{letltxmacro}
\LetLtxMacro{\oldcite}{\cite}
\renewcommand{\cite}[1]{\mbox{\oldcite{#1}}}

\newcommand{\be}{\begin{equation}}
\newcommand{\ee}{\end{equation}}
\newcommand{\bea}{\begin{eqnarray}}
\newcommand{\eea}{\end{eqnarray}}

\newcommand{\neff}{N_{\rm eff}}

\newcommand{\beq}{\begin{equation}}
\newcommand{\eeq}{\end{equation}}

\newcommand{\ps}[1]{{\it\color{blue}#1~\color{red}}}

\allowdisplaybreaks

\bibpunct{[}{]}{,}{n}{}{,}

\definecolor{darkgreen}{rgb}{0.0, 0.4, 0.0}
\begin{document}
\preprint{LAPTH-017/21}
\title{Massive sterile neutrinos in the early universe: From
thermal decoupling to cosmological constraints}

\author{Leonardo Mastrototaro}
\email{lmastrototaro@unisa.it}
\affiliation{Dipartimento di Fisica ``E.R Caianiello'', Università degli Studi di Salerno, Via Giovanni Paolo II, 132 - 84084 Fisciano (SA), Italy.}
\affiliation{Istituto Nazionale di Fisica Nucleare - Gruppo Collegato di Salerno - Sezione di Napoli, Via Giovanni Paolo II, 132 - 84084 Fisciano (SA), Italy.}

\author{Pasquale Dario Serpico}
\email{ serpico@lapth.cnrs.fr}
\affiliation{LAPTh,  Univ.   Grenoble  Alpes,  USMB,  CNRS,  F-74000  Annecy,  France.}

\author{Alessandro Mirizzi}
\email{alessandro.mirizzi@ba.infn.it }
\affiliation{Dipartimento Interateneo di Fisica ``Michelangelo Merlin'',
Via Amendola 173, 70126 Bari, Italy.}
\affiliation{Istituto Nazionale di Fisica Nucleare - Sezione di Bari,
Via Orabona 4, 70126 Bari, Italy.}

\author{Ninetta Saviano}
\email{nsaviano@na.infn.it}
\affiliation{INFN - Sezione di Napoli, Complesso Univ. Monte S. Angelo, I-80126 Napoli, Italy}
\affiliation{Scuola Superiore Meridionale, Università  degli studi di Napoli ``Federico II'', Largo San Marcellino 10, 80138 Napoli, Italy}

%=============================================================================

\begin{abstract}
We consider relatively heavy neutrinos  $\nu_H$, mostly contributing to a sterile state $\nu_s$,  with mass in the range 10 MeV $\lesssim m_s  \lesssim  m_{\pi} \sim 135$ MeV, which are thermally produced in the early universe in collisional processes involving active neutrinos, and freezing out after the QCD phase transition. If these neutrinos decay after the active neutrino decoupling, they generate extra neutrino radiation, but also contribute to entropy production. Thus, they alter the value of the effective number of neutrino species $N_{\rm eff}$ as for instance measured by the cosmic microwave background (CMB), as well as affect primordial nucleosynthesis (BBN), notably ${}^4$He production. We provide a detailed account of the solution of the relevant Boltzmann equations. We also identify the parameter space allowed by current 
Planck satellite data and forecast the parameter space probed by future Stage-4 ground-based CMB observations, expected to match or surpass BBN sensitivity.

  \end{abstract}

\maketitle

\section{Introduction}

Feebly interacting particles  characterised  by  extremely suppressed interactions with
the Standard Model  particles have received growing interest in the last decade 
(see~\cite{Agrawal:2021dbo} for a recent review).
In this context, a fourth neutrino mass state $\nu_H$ with mass $\sim {\mathcal O}(100)$ MeV, mostly contributing to an electroweak singlet neutrino state $\nu_s$ due to $Z$-width constraints~\cite{ALEPH:2005ab}, emerges rather naturally in extensions of the Standard Model,  like dynamical electroweak symmetry breaking~\cite{Appelquist:2002me}  or the Neutrino Minimal Standard Model ($\nu$MSM)~\cite{Asaka:2005an,Asaka:2005pn}.
In the latter case, such particles can be related to fundamental problems of particle physics like the origin of neutrino mass, the baryon asymmetry in the early universe and the nature of dark matter. 

The parameter space of a fourth neutrino in this mass range is strongly constrained by collider and beam-dump experiments for a dominant mixing with either $\nu_e$ and $\nu_\mu$~\cite{Alekhin:2015byh,Chun:2019nwi}, but it is significantly less constrained if mixed with $\nu_\tau$, with bounds at high masses coming from searches of decays of $D$ mesons and $\tau$ leptons~\cite{Orloff:2002de} and SuperKamiokande data~\cite{Coloma:2019htx}. 
Furthermore,  $\nu_H$  can be emitted by a core-collapse supernova.
 In this context, limits have been placed from the SN 1987A observation, requiring the SN core may not emit too much energy in the $\nu_s$ channel, since this additional energy-loss would shorten the observed neutrino burst \cite{Dolgov:2000pj,Dolgov:2000jw,Fuller:2009zz,Mastrototaro:2019vug}. 
Additionally, in \cite{Fuller:2009zz,Rembiasz:2018lok} it has been discussed a possible role of heavy sterile neutrinos in enhancing supernova explosions. 

Further and complementary constraints on heavy sterile neutrinos can also be placed from cosmological arguments.
Indeed,  $\nu_s$ can be produced in the early universe via collisional processes involving active neutrinos,
and then decay into lighter species.
In particular, for  masses in the range 10 MeV $\lesssim  m_s  \lesssim  m_{\pi} \sim 135$ MeV the main decay channels are 
$\nu_s \to \nu_\alpha\bar\nu_\beta\nu_\beta$ (with branching ratio of ${\cal O}(60-90\%)$, depending on the mixing) and $\nu_s \to \nu e^+ e^-$  (with branching ratio of ${\cal O}(10-40\%)$, depending on the mixing).
The decay products of the sterile neutrinos are injected into the primordial plasma, with the timescale of the event determining its phenomenological impact.
If the decay is over when active neutrinos are still tightly coupled to the electromagnetic (e.m.) plasma dominated by photons and $e^\pm$ pairs, full equilibrium conditions
are quickly established and no effect remains, but for an unobservable renormalisation of the baryon to photon ratio 
$\eta$~\footnote{This process leads instead to observable consequences if the renormalisation happens during or after primordial nucleosynthesis (BBN), creating e.g. an effective mismatch of $\eta_{\rm BBN}$ with respect to
the value extracted from Cosmic Microwave Background (CMB) $\eta_{\rm CMB}$. This is not relevant in the parameter space explored here, and we will ignore it in the following.}. 
If significant decay happens after the active neutrinos have decoupled, part of the energy injected will end up in extra neutrino radiation and part heats the e.m. plasma up. The latter adds to the eventual heating due to  $e^+ e^-$ annihilation and increases the photon to neutrino temperature beyond its standard value of $T_\gamma/T_\nu \simeq (11/4)^{1/3}$. Together with the former effect, this alters the effective number of neutrino species $N_{\rm eff}$, with the two processes going in opposite directions. Also, the non-thermal $\nu_e$ and $\bar{\nu}_e$ spectra enter weak interactions, altering---together with $N_{\rm eff}$---the \emph{neutron-to-proton ratio} which rules the abundance of the primordial yields \cite{Dolgov:2000pj,Dolgov:2000jw, Drewes:2015iva, Domcke:2020ety,Ruchayskiy:2012si,Sabti:2020yrt}, affecting  in particular the $^4$He abundance encoded in the primordial Helium mass fraction parameter $Y_p$.

The aim of our paper is to perform a detailed calculation of heavy sterile neutrino decoupling in the early universe, 
in particular computing the effects on  $N_{\mathrm{eff}}$ and  $Y_p$ and assessing the impact of the approximations presented in the seminal works~\cite{Dolgov:2000pj,Dolgov:2000jw}. 
Compared instead to the most recent calculations such as~\cite{Sabti:2020yrt,Boyarsky:2020dzc}, we put more emphasis on the low-mass sterile neutrino range, where the number
of effects is limited, hopefully offering a pedagogically complete and contained treatment, besides clarifying some points of disagreement. At high masses, new channels involving pions open up and
are responsible for extra effects on primordial nucleosynthesis as well as $\neff$ (some of these interesting effects have been described in~\cite{ Ruchayskiy:2012si,Sabti:2020yrt,Boyarsky:2020dzc}). In the same spirit, we will limit ourselves to neutrinos that decouple after the QCD phase transition,
which translates on the chosen parameter space.
One of our primary goals is to compare the effect on BBN with the latest constraints by the Planck satellite experiment as well as forecasts of future Stage-4 (S4) ground-based CMB observations~\cite{Abazajian:2019eic}. We shall illustrate the shifting cosmological constraining power, from a BBN-dominated one to a CMB-dominated one, expected to be basically completed by the S4 era.

The plan of our work is as follows. In Sec.~\ref{model} we present the heavy sterile neutrino model we will use as a benchmark. 
In Sec.~\ref{evolution} we discuss and solve the kinetic equations describing the sterile neutrino evolution in the early universe.
In Sec.~\ref{constraints} we characterise the impact of heavy sterile neutrino decays on active neutrinos and on $N_{\rm eff}$.
In Sec.~\ref{C&F} we present the current constraints and forecasts on sterile neutrino parameter space from BBN and CMB data.
Finally, in Sec.~\ref{summary} we summarise our results and conclude. In Appendix~\ref{appA} we report useful analytical approximations of the
decay and scattering rates involving sterile neutrinos. Appendix~\ref{appB} details the steps involved in the dimensional reduction of the collision integrals used in the numerical
integrations. Appendix~\ref{appC} is devoted to a comparison of our results with others previously reported in the literature.

%%%%%%%%%%%%%%
\section{Heavy sterile neutrino model}\label{model}
%%%%%%%%%%%%%%%

We consider heavy sterile neutrinos with masses 10 MeV $\lesssim m_s\lesssim $ 135 MeV, mixed dominantly with  one active neutrino 
$\nu_\alpha$ ($\alpha=e,\mu,\tau$)
as
%..........................
\begin{eqnarray}
\nu_\alpha &=& \cos \theta_{\alpha s} \nu_\ell + \sin\theta_{\alpha s} \nu_H \,\ ,  \nonumber \\ 
\nu_s &=& -\sin \theta_{\alpha s} \nu_\ell + \cos\theta_{\alpha s} \nu_H \,\ ,
\end{eqnarray}
%.........................
where $\nu_\ell$ and $\nu_H$ are a light and a heavy mass eigenstate, respectively,  and
$\theta_{\alpha s} \ll 1$, i.e.  $\nu_\ell$ is mostly active and $\nu_H$ is mostly sterile.
We can relate the  mixing angle to the unitary mixing matrix $U$,
where
%.....................
\begin{equation}
|U_{\alpha s}|^2 \simeq \frac{1}{4} \sin^2 2 \theta_{\alpha s}\simeq \theta_{\alpha s}^2 \,\ .
\end{equation}
%........................
Through neutral-current interactions, $\nu_H$ can decay into a $\nu_\ell$ and a pair of other
light leptons. It can also scatter with other species in the plasma.
With a little abuse of notation, we shall refer to $\nu_H$ as $\nu_s$ and to $\nu_\ell$ as $\nu_\alpha$.
Unless stated otherwise, we shall consider neutrinos to be Dirac particles.

In vacuum, in the mass range of our interest, the decay rate of sterile neutrinos is dominated by three lepton final states. In what follows,
we shall neglect terms of the order $m_\nu/m_{W,Z}$ in the matrix elements.  Their relevant decay processes and matrix elements are presented in Table~\ref{decay}.
Unless stated otherwise, we adopt the case of mixing with $\nu_\tau$ as our benchmark. In some cases, we will consider mixing with $\nu_e$ in the mass range  $m_s<m_e+m_\mu$ in order to compare with previous literature.~\footnote{This is just a simplification to avoid including the additional charged current decay channel $\nu_s\to e+\nu_\mu+\mu$.}

%%%%%%%%%%%%%%%%%%%%%%%%%%%%%%%%%%%%%%%%%%%%%%%%%%%%%%%%%%%%%%%%%%%%%%%%%%%%%%%%%%%%%%%%%%%%%%%%
\begin{table}
\centering
\caption{Squared matrix elements for sterile neutrino decay processes (assuming mixing with the species $\alpha$, and $\beta\neq\alpha$),  summed over initial and final states and divided by the spin dof of the sterile neutrino. The particles involved in each decay are enumerated as $1\rightarrow 2+3+4$. In the last lines,  $\tilde{g}_L $ is replaced by $g_L$ in case of mixing with $\nu_e$. The symmetry factor $S=1/2!$ is included,  when two identical particles are present in the final state (first row).}
\begin{tabular}{cc}
\hline
Process  &$G_F^{-2}|U_{\alpha s}|^{-2}|M|^2$\\
\hline
\hline
$\nu_s\rightarrow\nu_{\alpha}+\bar{\nu}_{\alpha}+\nu_{\alpha}$ &$32(p_1\cdot p_4)(p_2\cdot p_3)$ \\
$\nu_s\rightarrow\nu_{\alpha}+\nu_{\beta}+\bar{\nu}_{\beta}$ &$16(p_1\cdot p_4)(p_2\cdot p_3)$\\
$\nu_s\rightarrow\nu_{\alpha}+e^++e^-$ &$64[\tilde{g}_L^2(p_1\cdot p_4)(p_2\cdot p_3)+g^2_R(p_1\cdot p_3)(p_2\cdot p_4)-\tilde{g}_Lg_Rm_e^2(p_1\cdot p_3)]$\\
\hline
\end{tabular}
\label{decay}
\end{table}
%%%%%%%%%%%%%%%%%%%%%%%%%%%%%%%%%%%%%%%%%%%%%%%%%%%%%%%%%%%%%%%%%%%%%%%%%%%%%%%%%%%%%%%%%%%%%%%%%

The decay rate of sterile neutrinos into three neutrinos (summed over all flavours) is given by (see e.g.~\cite{Bondarenko:2018ptm})
%......................
\begin{equation}
\sum_\beta\Gamma(\nu_s\to \nu_\tau\bar{\nu}_\beta\nu_\beta)=\frac{|U_{\tau s}|^2}{192 \pi^3} G_F^2 m_{s}^5\label{3nu}\,,
\end{equation}
%......................
while the decay into neutrino plus $e^+e^-$ pair, in the limit where  $m_e/m_{s}\ll 1$ is neglected, is
%......................
\begin{equation}
\Gamma(\nu_s\to \nu_\tau e^+e^-)= \frac{|U_{\tau s}|^2}{192 \pi^3} G_F^2 m_{s}^5( \tilde{g}_L^2+g_R^2)\,,\label{nuee}
\end{equation}
%......................
where
%...........
\begin{eqnarray}
 g_L &=& \frac{1}{2} + \sin^2 \theta_W \,\ , \nonumber \\
 \tilde{g}_L &=& g_L -1= -\frac{1}{2} + \sin^2 \theta_W \,\ , \nonumber \\
 g_R &=&  \sin^2 \theta_W \,\ ,
\end{eqnarray}
%..............
and, in case of mixing with $\nu_e$, $ \tilde{g}_L $ is replaced by $g_L$.
As a result, the total decay width writes 
%......................
\begin{equation}
\Gamma_{\nu_s}= \tau_s^{-1}
= \frac{|U_{\tau s}|^2}{192 \pi^3} G_F^2 m_{s}^5 (1+\tilde{ g}_L^2+g_R^2)=3.90\times 10^{5}|U_{\tau s}|^2\left(\frac{m_s}{100~\mathrm{MeV}}\right)^5\mathrm{s^{-1}} \,\ .
\label{eq:decay}
\end{equation}
%......................

Note that for Majorana neutrinos, the widths for the exclusive decay processes would be the same as for the Dirac case, 
but the inclusive decay is a factor of two larger, since for each final state accessible to a Dirac particle, two channels are present (i.e. 
a channel and its $\sf L$-number conjugate). 

Two-body reactions affect the sterile neutrino chemical and kinetic equilibrium. Those of interest for us are reported, 
together with the relevant squared matrix elements, in Table~\ref{scattering}. Our results agree with those reported in 
table  5 and table 6 of Ref.~\cite{Sabti:2020yrt}, bearing in mind their different definition
of $|M|^2$, just summed over helicities of initial and final states, instead of averaged over 
the initial state as in our case.~\footnote{This is explicitly shown in their Eq.~(D.1), where their factor two larger value in $|M|^2$ is compensated by the further $1/g$  in their integral prefactor, $g=2$ being the spin multiplicity.} 
In relation instead to the Ref.~\cite{Dolgov:2000pj,Dolgov:1997mb}, we find systematically lower matrix elements compared to their Tables 1, 2, by a factor 2. We believe that this is likely a typo in the quantities reported rather than in their results since we do find agreement with the rates they compute.

%%%%%%%%%%%%%%%%%%%%%%%%%%%%%%%%%%%%%%%%%%%%%%%%%%%%%%%%%%%%%%%%%%%%%%%%%%%%%%%%%%%%%%%%%%%%%%%%
\begin{table}
\centering
\caption{Squared matrix elements for sterile neutrino scattering processes  (assuming mixing with the species $\alpha$, and $\beta\neq\alpha$), summed over initial and final states and divided by the two spin dof of the sterile neutrino. The particles involved in each reaction are enumerated as $1+2\rightarrow 3+4$. In the last line,  $\tilde{g}_L $ is replaced by $g_L$ in case of mixing with $\nu_e$. The symmetry factor $S=1/2!$ is included,  when two identical particles are present in the final state (second row).}
\begin{tabular}{cc}
\hline
Process  &$G_F^{-2}|U_{\tau s}|^{-2}|M|^2$\\
\hline
\hline
$\nu_s+\bar{\nu}_{\alpha}\rightarrow\nu_{\alpha}+\bar{\nu}_{\alpha}$ &$64(p_1\cdot p_4)(p_2\cdot p_3)$ \\
$\nu_s+\nu_{\alpha}\rightarrow\nu_{\alpha}+\nu_{\alpha}$ &$32(p_1\cdot p_2)(p_3\cdot p_4)$\\
$\nu_s+\bar{\nu}_{\alpha}\rightarrow\nu_{\beta}+\bar{\nu}_{\beta}$ &$16(p_1\cdot p_4)(p_2\cdot p_3)$\\
$\nu_s+\bar{\nu}_{\beta}\rightarrow\nu_{\alpha}+\bar{\nu}_{\beta}$ &$16(p_1\cdot p_4)(p_2\cdot p_3)$ \\
$\nu_s+\bar{\nu}_{\alpha}\rightarrow e^++e^-$ &$64[\tilde{g}_L^2(p_1\cdot p_4)(p_2\cdot p_3)+g^2_R(p_1\cdot p_3)(p_2\cdot p_4)-\tilde{g}_Lg_Rm_e^2(p_1\cdot p_3)]$\\
$\nu_s+e^-\rightarrow\nu_{\alpha}+e^-$ &$64[\tilde{g}_L^2(p_1\cdot p_2)(p_3\cdot p_4)+g^2_R(p_1\cdot p_4)(p_2\cdot p_3)-\tilde{g}_Lg_Rm_e^2(p_1\cdot p_3)]$\\
$\nu_s+e^+\rightarrow\nu_{\alpha}+e^+$ &$64[g_R^2(p_1\cdot p_2)(p_3\cdot p_4)+\tilde{g}^2_L(p_1\cdot p_4)(p_2\cdot p_3)-\tilde{g}_Lg_Rm_e^2(p_1\cdot p_3)]$\\
\hline
\end{tabular}
\label{scattering}
\end{table}
%%%%%%%%%%%%%%%%%%%%%%%%%%%%%%%%%%%%%%%%%%%%%%%%%%%%%%%%%%%%%%%%%%%%%%%%%%%%%%%%%%%%%%%%%%%%%%%%%

\section{Sterile and active neutrino evolution in the early universe}\label{evolution}

\subsection{Equations of motion}

Following \cite{Esposito:2000hi}, in order to describe the time evolution of the sterile neutrino ensemble in the early universe, it proves useful to define the following dimensionless variables which replace time, momentum and photon temperature, respectively
%...............
\begin{equation}
x\equiv m a \,\ \,\ \,\ \,\ \,\ \,\  y=p a \,\ \,\ \,\ z=T a \,\ ,
\label{eq:eom}
\end{equation}
%.............
where $m$ is an arbitrary mass scale which we set equal to $1$ MeV. Note that the function $a$ can be normalized, without loss of generality, so that
$ z=1$ at the largest temperature of interest here, when all particles in the plasma are in equilibrium with each other, and neutrinos also share the same temperature $T$.
In terms of these variables, we can write the equations of motion (EoMs) for the heavy sterile neutrinos distribution function $f_{\nu_s}$ as \cite{Dolgov:2000pj,Dolgov:2000jw}
%......................
\begin{equation}
H x \partial_x f_{\nu_s}= I_{\nu_s}[f_{\nu_s}]\,\ .
\label{eq:sterile}
\end{equation}
%...................
A similar equation holds for the evolution of  active neutrino species $f_{\nu_\alpha}$:
%...............
%......................
\begin{equation}
H x \partial_x f_{\nu_\alpha} = I_{\nu_\alpha}[f_{\nu_\alpha}] \,\ \,\ \,\ \,\ \nu_\alpha=\nu_e,\nu_\mu, \nu_\tau .
\label{eq:active}
\end{equation}
%...................
%.................
In the previous expressions, ${H}$ denotes 
the cosmic expansion Hubble rate  given by the Friedmann equation as
$H^2=8\pi \rho/(3 m_{\rm Pl}^2)$  where $\rho$ is the total energy density and 
$m_{\rm Pl}=G_{\rm N}^{-1/2}$ is the Planck mass in terms of the Newton constant $G_{\rm N}$. 
We will consider the plasma to be initially thermally populated by pions and all lighter particles,  
neglecting the nuclei contribution to $\rho$ and assuming equal distributions of particles and antiparticles.

The right-hand-side (r.h.s.) term in Eq.~(\ref{eq:sterile})--(\ref{eq:active}) contains the sum of the collisional and decay terms
for sterile and active neutrinos, respectively
 and reads
%%%%%%%%%%%%%%
\begin{equation}
I[f_\nu]=\frac{1}{2E}\int\prod_i\left(\frac{d^3p_i}{2E_i(2\pi)^3}\right)\prod_f\left(\frac{d^3p_f}{2E_f(2\pi)^3}\right)  (2\pi)^4 \delta^{(4)}\left(\sum_i p_i-\sum_fp_f\right)|M_{fi}|^2F(f_i,f_f) \,\ , 
\label{general_coll_integrala}
\end{equation}
%%%%%%%%%%%%%%%%%%%%%%%
with $|M_{fi}|^2$ the sum of the squared-matrix elements over initial and final states, divided by the spin multiplicity of the state of interest. In the case of $I[f_{\nu_s}]$, it  contains  decay  and scattering processes,
as shown in Table \ref{decay} and \ref{scattering}, respectively. In the case of $I[f_{\nu_a}]$, it  contains  scattering processes analogous to 
those of Table~\ref{scattering}, apart for the replacement $m_s \to 0$ and $|U_{\tau s}|^2\to 1-|U_{\tau s}|^2$ or $1$, depending if one is dealing with the mixed flavour or not; it also contains  the  sterile neutrino decay source term, which is calculated referring to the processes in Table \ref{decay}.

We use squared-matrix elements for the collisional processes for active neutrinos consistent e.g. with the results of Table 3 of~\cite{Sabti:2020yrt}, modulo our different definition
of $|M_{fi}|^2$, while we find once again that the results reported e.g. in \cite{Dolgov:2000pj,Dolgov:1997mb} are a factor two larger than the correct ones.
The statistical factor writes in general as
%%%%%%%%%%%%%
\begin{equation}
F(f_i,f_f)=-\prod_if_i\prod_f(1\mp f_f)+\prod_i(1\mp f_i)\prod_ff_f \,\ ,
\label{Ffactor}
\end{equation}
%%%%%%%%%%%%%%%%%%%%%%
where $f_{i,f}$ are the distributions of the particles in the initial ($i$) or final ($f$) states, the  (-) sign (``blocking'') refers to fermions and (+) sign applies to bosons (``stimulated'' effect). Only fermions are however present in the processes of interest for us. 

Due to the different timescales of the active neutrino oscillation processes (see e.g.~\cite{Mangano:2005cc}), we take into account oscillations among active neutrinos by  post-processing the flavour distributions via 
%...................
\begin{equation}
I[f_\alpha] \to \sum_\beta P_{\alpha \beta} I[ f_\beta]\,,
\end{equation}
%...................
where $P_{\alpha \beta}$ are the time-averaged transition probabilities (see e.g. Eq.~(26) in~\cite{Ruchayskiy:2012si}).  Medium modification of the mixing parameters
are also included, similarly to ref.~\cite{Sabti:2020yrt}.

To get the time evolution of sterile and active neutrino distributions, we have to complete the set of equations 
[Eqs.~(\ref{eq:sterile})--(\ref{eq:active})--(\ref{eq:hubble})] with the continuity equation, stating the conservation of the total energy density
%...........................
\begin{equation}
\frac{d}{dx}{\bar\rho}(x)=\frac{1}{x}({\bar \rho}-3 {\bar P}) \,\ ,
\label{eq:contin}
\end{equation}
%.............................
where ${\bar \rho}$ and ${\bar P}$ are the  comoving energy density and pressure of the primordial plasma,
respectively
%...................
\begin{eqnarray}
{\bar \rho} &=&  \rho \left(\frac{x}{m} \right)^4 \,\ , \nonumber \\
{\bar P} &=&  P \left(\frac{x}{m} \right)^4 \,\ .
\end{eqnarray}
%...................
Note that only massive components in the plasma contribute to the r.h.s. of Eq.~(\ref{eq:contin}), while (neglecting plasma corrections) relativistic species at equilibrium cancel out. 
From Eq.~(\ref{eq:contin}) by using the expressions in~\cite{Esposito:2000hi}, and remembering that we consider equal particle and antiparticle distributions,
%.............
\begin{eqnarray}
{\bar \rho}_{\gamma} &=& \frac{\pi^2}{15} z^4 \,\ , \\
{\bar \rho}_{\ell} &=& \frac{2}{\pi^2} \sum_\ell\int_{0}^\infty dy \,\  y^2 \frac{\sqrt{m_\ell^2x^2/m^2+y^2}}{\exp(\sqrt{m_\ell^2x^2/m^2+y^2}/z)+1} \,\ , \\
{\bar P}_\ell &=& \frac{2}{3 \pi^2} \sum_\ell\int_{0}^\infty dy \,\  \frac{y^4}{\sqrt{m_\ell^2x^2/m^2+y^2}}  \frac{1}{\exp(\sqrt{m_\ell^2x^2/m^2+y^2}/z)+1} \,\ , \\
{\bar \rho}_{\pi} &=& \frac{1}{2\pi^2} \sum_i\int_{0}^\infty dy \,\  y^2 \frac{\sqrt{m_{i}^2x^2/m^2+y^2}}{\exp(\sqrt{m_{i}^2x^2/m^2+y^2}/z)-1} \,\ , \\
{\bar P}_\pi &=& \frac{1}{6 \pi^2}\sum_{i} \int_{0}^\infty dy \,\  \frac{y^4}{\sqrt{m_{i}^2x^2/m^2+y^2}}  \frac{1}{\exp(\sqrt{m_{i}^2x^2/m^2+y^2}/z)-1} \,\ , \\
{\bar \rho}_{\nu_a} &=& 3 {\bar P}_{\nu_a}  =\frac{1}{\pi^2} \int_{0}^\infty dy \,\  y^3 \sum_\alpha f_{\nu_\alpha}(x,y) \,\ , \\
{\bar \rho}_{\nu_s} &=&  \frac{1}{\pi^2}\int_{0}^\infty dy \,\ y^2\sqrt{m_s^2x^2/m^2+y^2} \,\  f_{\nu_s}(x,y) \,\ ,\\
{\bar P}_{\nu_s} &=&  \frac{1}{3\pi^2}\int_{0}^\infty dy \,\ \frac{y^4}{\sqrt{m_s^2x^2/m^2+y^2}} \,\  f_{\nu_s}(x,y) \,\ ,
\end{eqnarray}
%...............
where $i$ runs over the three pions, $\pi^+,\pi^-,\pi^0$,  and $\ell$ is either
the muon or the electron.
Note that the Hubble function can be now expressed as
\begin{equation}
H
=\sqrt{\frac{8\pi}{3}}\,
\frac{1}{m_{\rm Pl}}\frac{m^2}{x^2}\left({\bar \rho}_{\gamma}+{\bar \rho}_{e}+{\bar \rho}_{\mu}+{\bar \rho}_{\pi}+ {\bar \rho}_{\nu_a}+{\bar \rho}_{\nu_s} \right)^{1/2}\,.
\label{eq:hubble}
\end{equation}
Equation~(\ref{eq:contin}) gets contributions from all species, and can be recast into the equation for the $z(x)$ relation. Let us specify the different contributions.
For photons one  has:
\begin{equation}
\frac{{\rm d}\bar{\rho}_{\gamma}}{{\rm d}x}-\frac{\bar{\rho}_\gamma-3\bar{P}_\gamma}{x}=\frac{{\rm d}\bar{\rho}_{\gamma}}{{\rm d}x}= \frac{4\pi^2\, z^3}{15}\frac{{\rm d}z}{{\rm d}x}.
\end{equation}

For the electrons,  if setting $\kappa_e\equiv m_e/m$, 
one finds:
\begin{equation}
\frac{{\rm d}\bar{\rho}_e}{{\rm d}x}-\frac{\bar{\rho}_e-3\bar{P}_e}{x}=
\frac{2z^3}{\pi^2}\left\{-\kappa_e^2\frac{x}{z}F_1^+\left(\frac{\kappa_ex}{z}\right)+\frac{{\rm d}z}{{\rm d}x}\left[\kappa_e^2\frac{x^2}{z^2}F_1^+\left(\frac{\kappa_e\,x}{z}\right)+F_2^+\left(\frac{\kappa_e\,x}{z}\right)\right]\right\}
  \,\ ,
\end{equation}
where the functions $F_1^\pm$ and $F_2^\pm$ are defined as
%......................................
\begin{eqnarray}
F_1^\pm(\tau)&\equiv&\int_0^{\infty}d\omega\,\omega^2\frac{\exp(\sqrt{\omega^2+\tau^2})}{(\exp(\sqrt{\omega^2+\tau^2})\pm1)^2} \,\ , \nonumber \\
F_2^\pm(\tau)&\equiv&\int_0^{\infty}d\omega\,\omega^4\frac{\exp(\sqrt{\omega^2+\tau^2})}{(\exp(\sqrt{\omega^2+\tau^2})\pm1)^2} \,\ ,
\end{eqnarray}
with the signs $\pm$ that take into account the boson and fermion nature of the particle respectively.
A similar expression holds for muons, with $\kappa_e\to \kappa_\mu$.
For each pion species $i$, one has instead:
\begin{equation}
\frac{{\rm d}\bar{\rho}_i}{{\rm d}x}-\frac{\bar{\rho}_i-3\bar{P}_i}{x}=
\frac{z^3}{2\pi^2}\left\{-\kappa_i^2\frac{x}{z}F_1^-\left(\frac{\kappa_i\,x}{z}\right)+\frac{{\rm d}z}{{\rm d}x}\left[\kappa_i^2\frac{x^2}{z^2}F_1^-\left(\frac{\kappa_i\,x}{z}\right)+F_2^-\left(\frac{\kappa_e\,x}{z}\right)\right]\right\}
  \,\ .
\end{equation}

For sterile neutrinos ($\kappa_s\equiv m_s/m$):
\begin{equation}
\begin{split}
&\frac{{\rm d}\bar{\rho}_s}{{\rm d}x}-\frac{\bar{\rho}_s-3\bar{P}_s}{x}=\frac{1}{\pi^2}\int_{0}^\infty {\rm d}y \,\ y^2\left[\sqrt{\kappa_s^2x^2+y^2} \,\  \frac{\partial f_{\nu_s}(x,y)}{\partial x} +f_{\nu_s}(x,y)\left(\frac{\kappa_s^2x+y^2/x-(\kappa_s^2x+y^2/x)}{\sqrt{\kappa_s^2x^2+y^2}}\right)\right]=\\
&=\frac{1}{\pi^2}\int_{0}^\infty {\rm d}y \,\ y^2\sqrt{\kappa_s^2x^2+y^2} \,\  \frac{ \partial f_{\nu_s}(x,y)}{\partial x} \,\ .
\end{split}
\end{equation}

For active quasi-massless neutrinos, if $\alpha=e,\mu,\tau$, one has
\begin{equation}
\frac{{\rm d}\bar{\rho}_{\nu_a}}{{\rm d}x}-\frac{\bar{\rho}_{\nu_a}-3\bar{P}_{\nu_a}}{x}=\frac{{\rm d}\bar{\rho}_{\nu_a}}{{\rm d}x}=\sum_{\alpha}\frac{1}{\pi^2} \int_{0}^\infty {\rm d}y \, y^3  \frac{\partial f_{\nu_\alpha}}{\partial x}\,\ .
\end{equation}
Down to a few MeV, the active neutrinos are coupled to the rest of the plasma, which means that at sufficiently high temperatures (low $x$) we can write $f_{\nu_{\alpha}}=1/(\exp(y/z)+1)$ and 
\begin{equation}
\frac{\partial f_{\nu_\alpha}}{\partial x}=\frac{{\rm d}z}{{\rm d}x}\frac{\partial f_{\nu_{\alpha}}}{\partial z}=\frac{{\rm d}z}{{\rm d}x}\frac{y\exp(y/z)}{z^2(\exp(y/z)+1)^2}\:\:\:({\rm early}\:{\rm times})\,.\label{earlytimes}
\end{equation}

\begin{figure}
\centering
\includegraphics[scale=0.6]{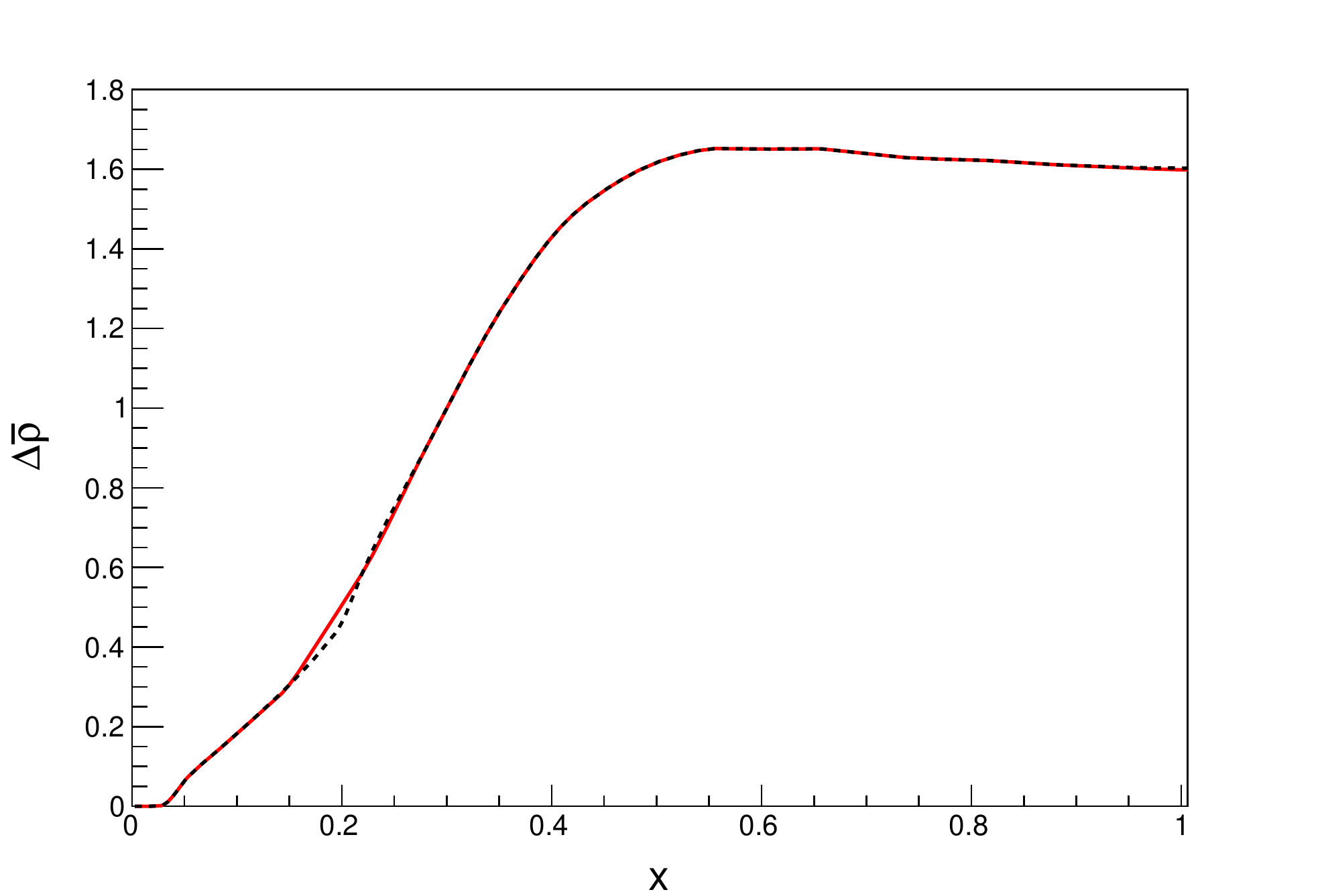}
\caption{Comparison between the comoving extra active neutrino energy using $x_d=0.1$ (red curve) and $x_d=0.2$ (black curve), for parameters $m_s=100~\mathrm{MeV}$ and $\tau_s=0.045~\mathrm{s}$.}
\label{01}
\end{figure}

As a result, in computing the $z=z(x)$ relation we can save considerable computer time by considering two different regimes: Eq.~(\ref{earlytimes}) for $x<x_d$, while numerically computing  $f_{\nu_\alpha}$ from the Boltzmann Eq.~(\ref{eq:active}) for $x>x_d$, where $x_d$ represents any epoch before neutrino decoupling, but otherwise arbitrary.
In terms of the step function $\Theta$,  collecting all terms for photons, electrons, pions, active and sterile neutrinos  and isolating ${\rm d}z/{\rm d}x$, Eq.~(\ref{eq:contin})
 can be written as:
\begin{equation}
\begin{split}
&\left[z^3\left(\frac{4\pi^2}{15}+\frac{A(x/z)}{\pi^2}\right)+\frac{3\Theta(x_d-x)}{\pi^2z^2}\int_{0}^\infty {\rm d}y \, y^4
\frac{\exp(y/z)}{(\exp(y/z)+1)^2}\right]\frac{{\rm d}z}{{\rm d}x}=\\
&=\frac{z^3}{\pi^2}B(x/z)-\frac{1}{\pi^2}\int_{0}^\infty {\rm d}y \,\ y^2\left(\sqrt{\kappa_s^2x^2+y^2} \,\  \frac{\partial f_{\nu_s}(x,y)}{\partial x}+y\,\Theta(x-x_d)\sum_\alpha\frac{\partial f_{\nu_{\alpha}}}{\partial x}\right) \,\ ,
\end{split}
\label{temperature evolution}
\end{equation}
where we defined:
\begin{equation}
A(w)=2\sum_\ell\left(\kappa_\ell^2w^2F_1^+\left(\kappa_\ell w\right)+F_2^+\left(\kappa_\ell\,w\right)\right)+\frac{1}{2}\sum_i\left(\kappa_i^2w^2 F_1^-\left(\kappa_i\,w\right)+F_2^-\left(\kappa_i\,w\right)\right)\,,
\end{equation}
and
\begin{equation}
B(w)=2\sum_\ell\kappa_\ell^2\,w F_1^+\left(\kappa_\ell \, w\right)+\frac{1}{2}\sum_i\kappa_i^2w F_1^-\left(\kappa_i\,w\right)\,.
\end{equation}
Together with   $z=1$ as initial condition, Eq.~(\ref{temperature evolution}) gives  the ``time-temperature'' evolution. Provided that $x_d$ is sufficiently small, roughly $x_d\lesssim 0.2$ (i.e. $T\gtrsim 5\,$MeV), the computed behaviour is insensitive to the choice of $x_d$, as we illustrate in Fig.~\ref{01}, where  the extra comoving neutrino energy density evolution  is computed using $x_d=0.1$ (red solid curve) and $x_d=0.2$ (black dotted curve), for parameters $m_s=100~\mathrm{MeV}$ and $\tau_s=0.045~\mathrm{s}$. The results are almost equal except for small numerical differences when $0.1<x<0.2$. In the following, we fix $x_d=0.1$.
Note that the parenthesis at the r.h.s. of Eq.~(\ref{temperature evolution}) describes the ``heating'' of the e.m. plasma due to the sterile neutrino entropy release: At early times, all of the decay products end up in the coupled plasma of photons and active neutrinos, raising $z$ compared to standard expectations.  At late times, a sizable fraction
decays into (decoupled) active neutrinos, hence the term in parenthesis largely cancels out.  The only non-trivial evolution of $z$ is then due to the finite mass term of $e^\pm$, affecting their annihilation at late time. 
%%%%%%%%%%%%%%%%%%%%%%%%%%%%%%%%%%
%%%%%%%%%%%%%%%%%%%%%%%%%%%%%%%%%%

%%%%%%%%%%%%%%%%
\subsection{Evolution of heavy sterile neutrinos}
%%%%%%%%%%%%%%%%

In~\cite{Dolgov:2000jw}, an analytical solution of Eq.~(\ref{eq:sterile}) was provided under the following assumptions:
%%%%%%%%%%%%
\begin{itemize}
\begin{comment}
\item[\emph{(i)}] Sterile neutrinos are assumed relativistic, with an energy-momentum relation approximated as $E\simeq p(1+0.5 m^2/p^2)$ and terms suppressed by higher powers of $m^2/p^2$ being neglected. 
 \ps{As far as I can read, both in~\cite{Dolgov:2000jw} and in~\cite{Dolgov:2000pj} state
that they consider {\it relativistic} sterile neutrinos at decoupling. }
\end{comment}
\item[\emph{(i)}] The equilibrium distribution functions ``inside'' the collisional integral [Eq.~(\ref{general_coll_integrala})]  are taken in the Boltzmann (classical) approximation, with Pauli blocking factors correspondingly neglected.
\item[\emph{(ii)}] Electrons are considered ultrarelativistic, with terms in $m_e^2$ neglected throughout.
\end{itemize}
%%%%%%%%%%
We repeated the derivation of~\cite{Dolgov:2000jw} under these approximation, finding:
\begin{equation}
I_{\mathrm{dec}}=\frac{(1+\tilde{g}_L^2+g_R^2)G_F^2m_1^5|U_{s\tau}|^2}{192\pi^3}\frac{m_s}{E_s}(f_s^{eq}-f_s) \equiv\frac{m_s}{E_s}\frac{1}{\tau_s}(f_s^{eq}-f_s) \,\ ,
\end{equation}
and
\begin{eqnarray}
I_{\mathrm{scatt}}&=&\frac{G_F^2|U_{s\tau}|^2(1+\tilde{g}_L^2+g_R^2)T^3m_s^2}{\pi^3}\left(f^{\mathrm{eq}}(E_1)-f(E_1)\right)\left[ \frac{3}{2}\zeta(3)+\frac{7T\pi^4}{72}\left(\frac{E_1}{m_s^2}+\frac{p_1^2}{3E_1m_s^2}\right)\right] \,\ \nonumber \\
&=&\frac{3\times 2^6}{\tau_s}\left(f^{\mathrm{eq}}(E_1)-f(E_1)\right)\left[\frac{3\zeta(3)}{2} \frac{T^3}{m_s^3}+\frac{7\pi^4}{72}\frac{T^4E_1}{m_s^5}\left(1+\frac{p_1^2}{3E_1^2}\right)\right],\label{finale1}
\end{eqnarray}
in agreement with their results quoted as 
%............
\begin{equation}
 \partial_xf_{\nu_s}(x,y)=\frac{1.48x}{\tau_s/s}\left(\sqrt{\frac{g_*}{10.75}}\right)^{1/2}
 \frac{\Bigg(f^{\mathrm{eq}}(E_1)-f(E_1)\Bigg)}{(xT)^2}\Bigg\lbrace\frac{m_s}{E_s}+3\times 2^7T^3\Bigg[\frac{3}{4}\frac{\zeta(3)}{m_s^3}+\frac{7T\pi^4}{144}\Bigg(\frac{E_1}{m_s^5}+\frac{p_1^2}{3E_1m_s^5}\Bigg) \Bigg]\Bigg\rbrace \,\ ,
\label{finale}
\end{equation}
%..................
where $f_{\nu_s}^{\mathrm{eq}}$ is the Fermi-Dirac equilibrium distribution of the sterile neutrinos, 
$\tau_s$ is the sterile neutrino lifetime, and $E_s=\sqrt{m_s^2 +(y/x)^2}$ is the sterile neutrino energy.
Details of the reduction of integrals in Eq.~(\ref{general_coll_integrala}) under approximations $(i)$ and $(ii)$ are given in the Appendix~\ref{appA}.

We also solve the sterile neutrino kinetic equations numerically, relaxing the approximations $(i)$ and $(ii)$ in Eq.~(\ref{general_coll_integrala}), for both sterile neutrinos and active neutrinos.  Following the well-known technique developed in~\cite{Hannestad:1995rs}, it is possible to  analytically reduce the nine-dimensional 
collisional integral into a two-dimensional one, which is then integrated numerically. We developed an equivalent technique for the decay processes. We report the details in Appendix~\ref{appB}.

Our solutions are obtained by assuming initial thermal equilibrium for all species, starting from a temperature $T=\mathrm{min}[2\,m_s,150\,{\rm MeV}]$. To compare the difference in using numerical results vs. the analytical approximation, in Table~\ref{TD} we report the sterile neutrino freeze-out temperature $T_D$,  for a few representative points in parameter space, according to the condition $I(T_D)=H(T_D)$, with $I$ given by 
Eq.~(\ref{general_coll_integrala}). We find typical differences at a few percent level, and in all cases below $10\%$.  Although we are using the numerical results in the following, when requiring only moderate precision on the sterile decoupling, the analytical approximation seems largely sufficient, and allows one to significantly gain in computing time.

%%%%%%%%%%%%%%%%%%%%%

%...........................
\begin{table}
\caption{Altered cosmologies in presence of sterile neutrinos. $\tau$ is the lifetime of the sterile neutrino considered, $T_D^a$ is the decoupling temperature obtained with the evolution in Eq.~(\ref{finale}), $T_D^n$ is the decoupling temperature obtained solving the Boltzmann equation numerically, and $Y_p$ the estimated value of $^4$He abundance, discussed in sec.~\ref{impactYp}.}
\vspace{.5cm}
\centering
\begin{tabular}{cccccc}
\hline
$m_s~[\mathrm{MeV}]$&  $\sin^2\theta_{\tau 4}$& $\tau~[\mathrm{s}]$& $T^n_D~[\mathrm{MeV}]$& $T^a_D~[\mathrm{MeV}]$& $Y_p$\\
\hline
$20.0$& $2.6\times10^{-2}$& $3.0\times 10^{-1}$& $4.35$& $4.26$ & $0.2514$\\
$40.0$& $2.8\times10^{-3}$& $8.8\times 10^{-2}$& $9.24$& $10.00$ & $0.2520$\\
$60.0$& $5.5\times10^{-4}$& $6.0\times 10^{-2}$& $16.83$& $16.20$ & $0.2509$\\
$80.0$& $1.5\times10^{-4}$& $5.0\times 10^{-2}$& $26.53$& $25.22$ & $0.2628$\\
$100.0$& $5.8\times10^{-5}$& $4.4\times 10^{-2}$& $37.10$& $37.65$ & $0.2705$\\
$130.0$& $1.6\times10^{-5}$& $4.2\times 10^{-2}$& $59.13$& $59.00$ & $0.2881$\\
\hline
\end{tabular}
\label{TD}
\end{table}
%..................................

%%%%%%%%%%%%%%%%%%%%%%%%%%
\section{Impact on cosmological observables}

%%%%%%%%%%%%%%%%%%%%%%
\label{constraints}
%%%%%%%%%%%%%%%%%%%%%%%%
After the distribution functions and temperature evolution are found, we relate them to the observables  $N_{\mathrm{eff}}$ (notably at the CMB epoch) and $Y_p$ (notably at the BBN epoch) to derive some constraints.
%%%%%%%%%
\begin{figure}
\centering
\includegraphics[scale=0.6]{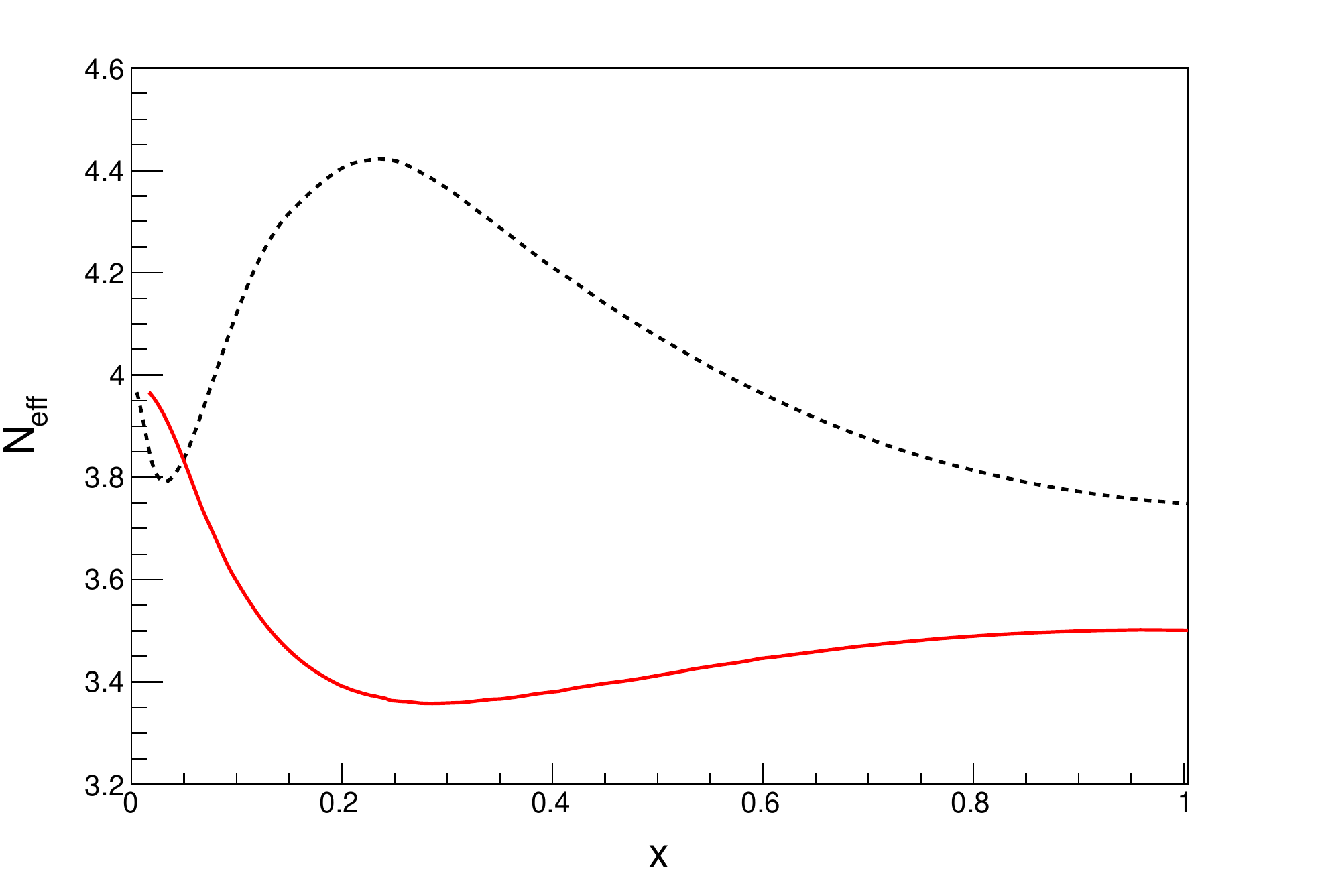}
\caption{$N_{\mathrm{eff}}$ evolution in $x$ for $m_s=30~\mathrm{MeV}$, $\tau_{\tau s}=0.15~\mathrm{s}$ (red, solid) and $m_s=100~\mathrm{MeV}$, $\tau_{\tau s}=0.055~\mathrm{s}$ (black, dashed).}
\label{Neff_100vs30}
\end{figure}
%%%%%%%

%%%%%%%%%
\begin{figure}
\centering
\includegraphics[scale=0.6]{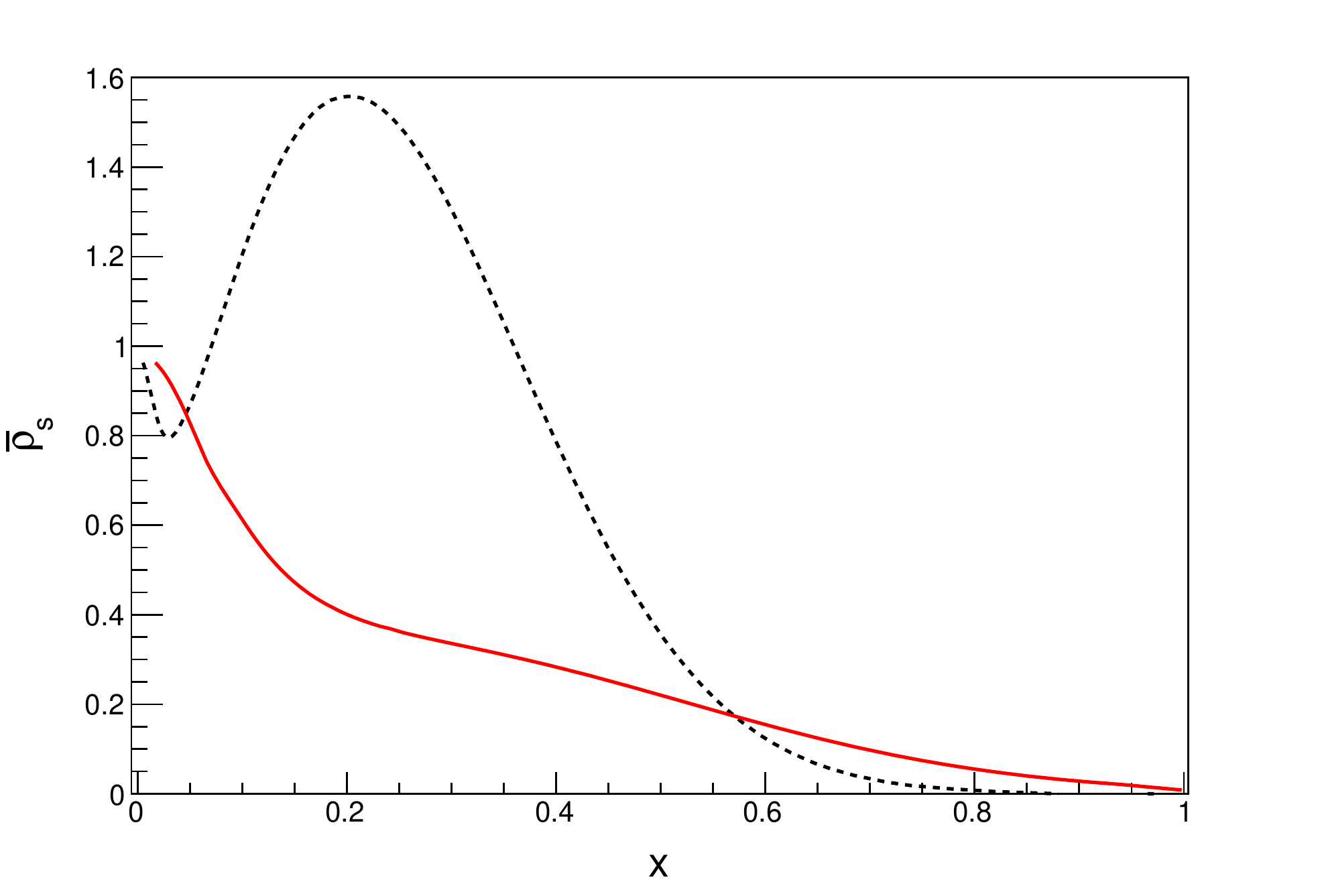}
\caption{Comoving energy density evolution for the sterile neutrino vs. $x$ for $m_s=30~\mathrm{MeV}$, $\tau_{\tau s}=0.15~\mathrm{s}$ (red, solid) and $m_s=100~\mathrm{MeV}$, $\tau_{\tau s}=0.055~\mathrm{s}$ (black, dashed).}
\label{rhos_100vs30}
\end{figure}
%%%%%%%

%%%%%%%%%
\begin{figure}
\centering
\includegraphics[scale=0.6]{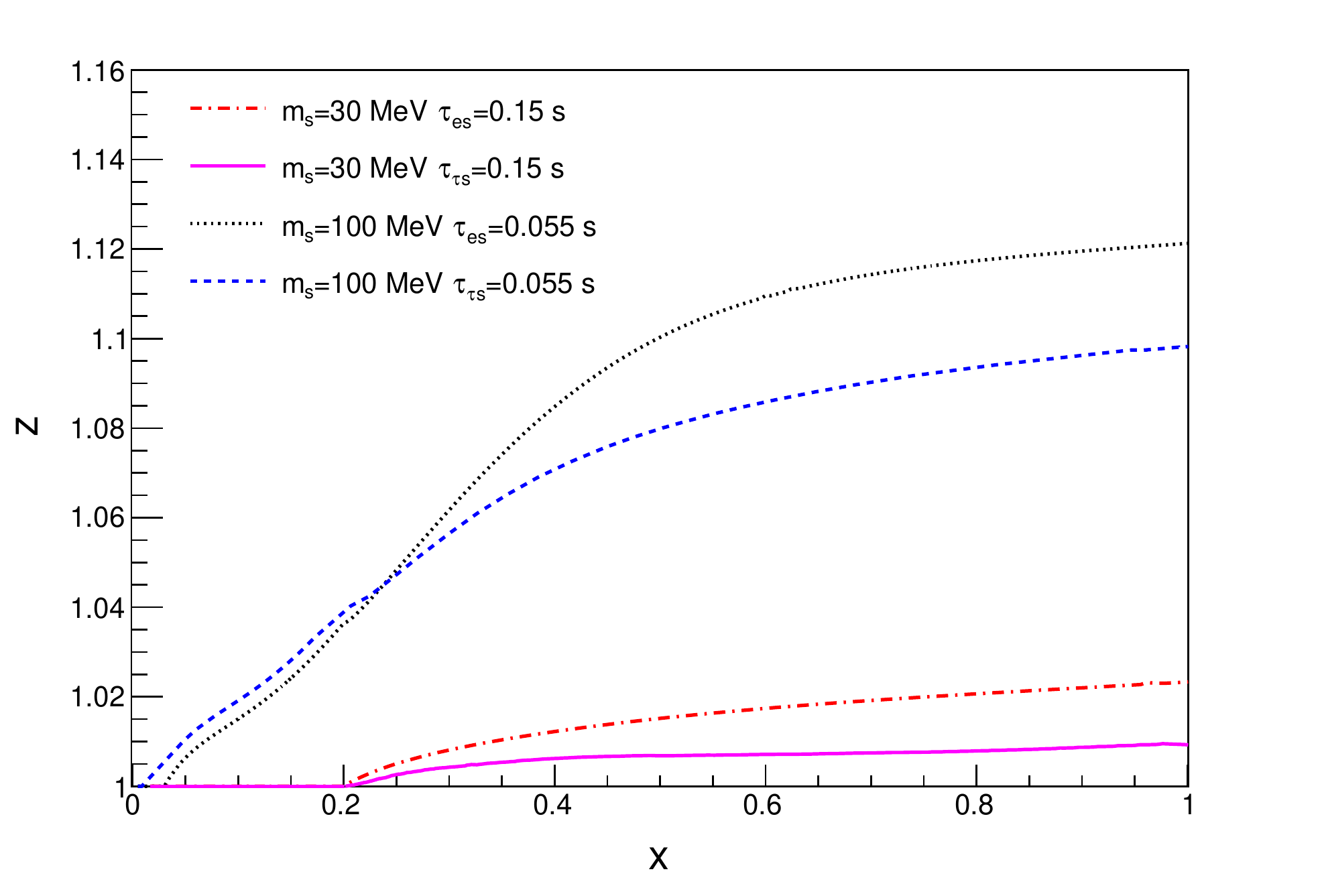}
\caption{Evolution of the dimensionless temperature $z$ vs. $x$ for $m_s=30~\mathrm{MeV}$, $\tau_{e s}=0.15~\mathrm{s}$ (red, dashed-dotted) $\tau_{\tau s}=0.15~\mathrm{s}$ (purple, solid) and $m_s=100~\mathrm{MeV}$, $\tau_{e s}=0.055~\mathrm{s}$ (blue, dashed) $\tau_{\tau s}=0.055~\mathrm{s}$ (black, dotted).}
\label{z_100vs30}
\end{figure}

%%%%%%%%%
\begin{figure}
\centering
\includegraphics[scale=0.6]{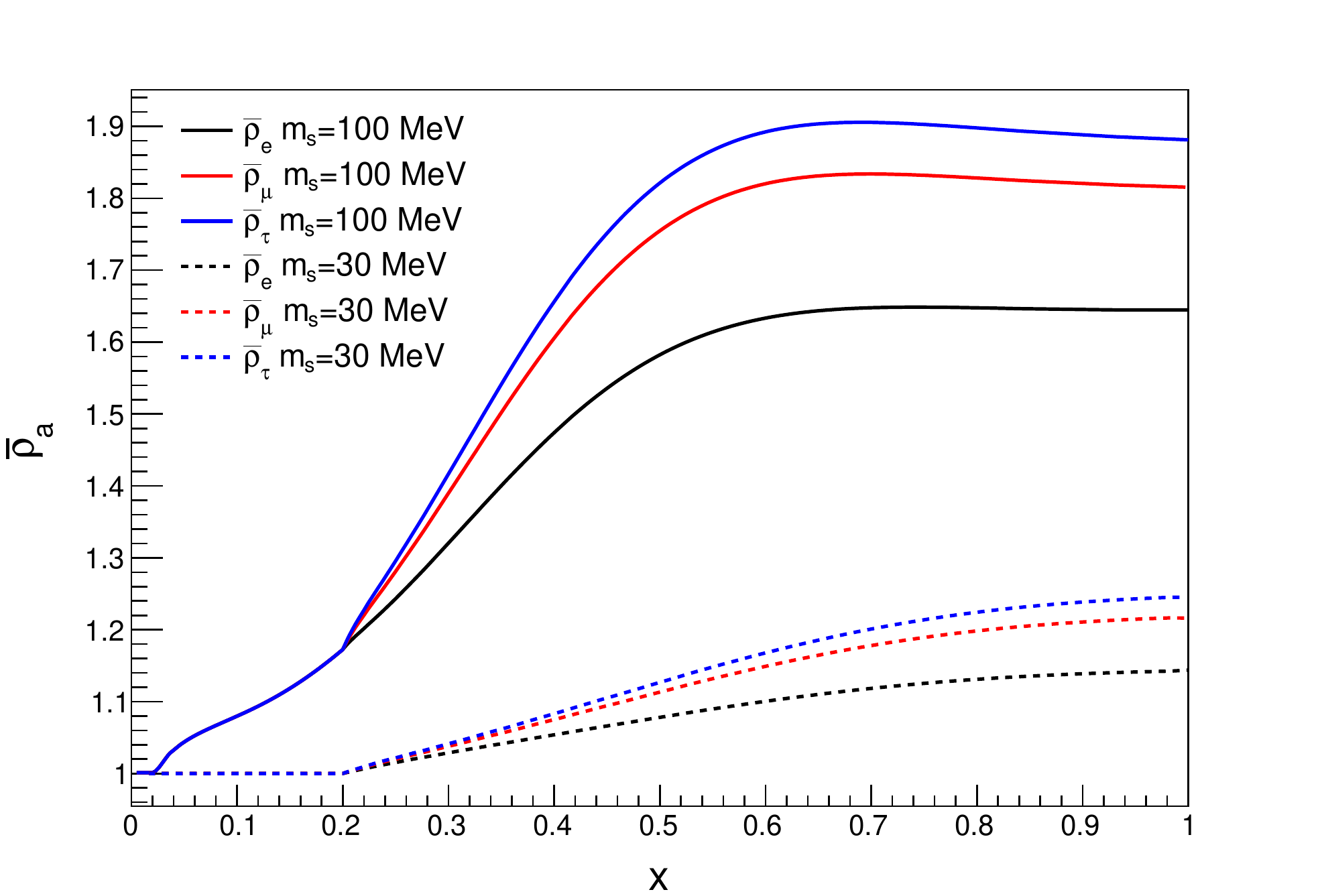}
\caption{Comoving energy density evolution for the three active neutrinos vs. $x$ for $m_s=30~\mathrm{MeV}$, $\tau_{\tau s}=0.15~\mathrm{s}$ (dashed lines) and $m_s=100~\mathrm{MeV}$, $\tau_{\tau s}=0.055~\mathrm{s}$ (solid lines). Each sets of curves, from top to bottom, represents the $\nu_\tau,\, \nu_\mu$ and $\nu_e$ energy density
evolution, respectively, reflecting the assumed sterile mixing with $\nu_\tau$ and the active neutrino mixing matrix.Here, $x_d=0.2$ is assumed.}
\label{rho_100vs30}
\end{figure}
%%%%%%%

%%%%%%%%%
\begin{figure}
\centering
\includegraphics[scale=0.6]{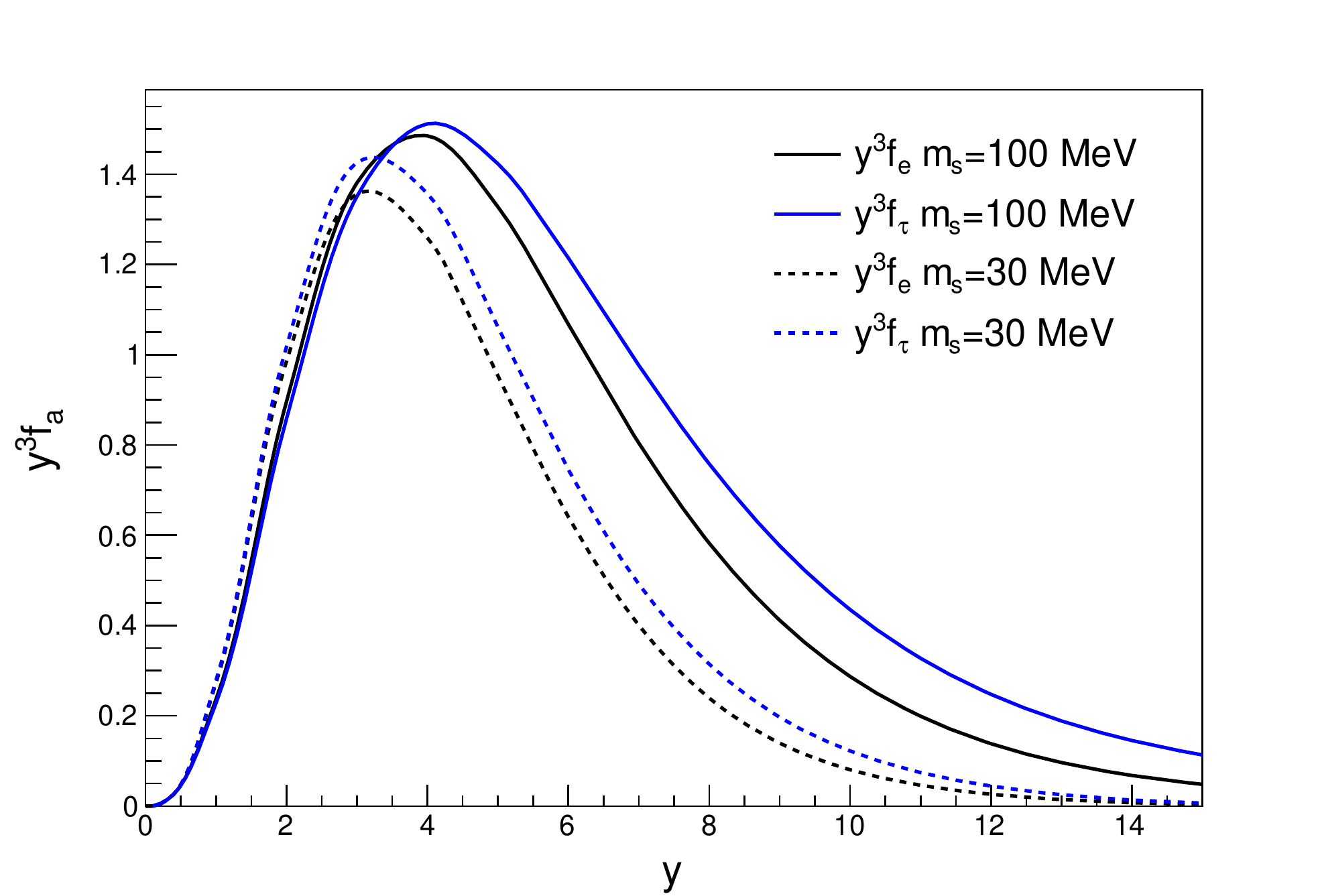}
\caption{$\nu_\tau$ and $\nu_e$ distribution functions vs. $y$ at $x=1$ for the same cases reported in Fig.~\ref{rho_100vs30}.}
\label{distrib}
\end{figure}
%%%%%%%
%%%%%%%%%%%%%%%%%%%%%%
\subsection{Impact on effective number of active neutrinos $N_{\rm eff}$}
%%%%%%%%%%%%%%%%%%%%%%%%
Heavy $\nu_s$ affect the total energy density in non-electromagnetic species. This is usually quantified in terms of the \emph{effective number of neutrinos}, $N_{\rm eff}$, which  is defined from the density in all species but electromagnetically interacting ones, as (see for instance~\cite{Mangano:2005cc})
\begin{equation}
N_{\rm eff}(x)=  \frac{\rho_\gamma^{\rm inst}}{\rho_\gamma}\sum_{i\neq \rm e.m.}\frac{\rho_i}{\rho_{\nu_0}}= \left(\frac{z_0(x)}{z(x)} \right)^4 \left(3+ \frac{\Delta \rho_{\nu_e}}{\rho_{\nu_0}} + 
\frac{\Delta \rho_{\nu_\mu}}{\rho_{\nu_0}}+\frac{\Delta \rho_{\nu_\tau}}{\rho_{\nu_0}} + \frac{\rho_{\nu_s}}{\rho_{\nu_0}} \right) \,, 
\label{eq:Neff}
\end{equation}
%................................
where the r.h.s. is specific for our 4 neutrino model, where $\Delta \rho_{\nu_\alpha}$ are the changes in the neutrino energy densities with respect to $\rho_{\nu_0}$, the energy density in the instantaneous decoupling limit, due to the non-equilibrium effects. Note that at early times around $T\simeq 100\,$MeV when all species are relativistic and share the same temperature, $N_{\rm eff}\to 4$. Asymptotically, when all sterile neutrinos have disappeared and the $e^+e^-$ annihilation is complete: i) $z_0\to (11/4)^{1/3} \simeq 1.4$, the asymptotic Standard Model photon-neutrino temperature ratio in the instantaneous decoupling limit; ii)  $z\to z_{\rm fin}$, the actual final photon/neutrino temperature; iii) $\rho_{\nu_s}/\rho_{\nu_0}\to 0$, since all sterile neutrinos have decayed away. At large $x$, we expect $N_{\rm eff}\gtrsim 3$ since, due to the branching fraction, the contribution due to extra radiation in the neutrino sector largely compensates the entropy transfer, which would tend to lower $N_{\rm eff}$ below 3 via the $z$-dependent pre-factor at the r.h.s. of Eq.~(\ref{eq:Neff}). Indeed this behavior can be seen in Fig.~\ref{Neff_100vs30} where we have plotted the $N_{\mathrm{eff}}$ evolution for $m_s=30~\mathrm{MeV}$, $\tau_{\tau s}=0.15~\mathrm{s}$ (in solid red) and $m_s=100~\mathrm{MeV}$, $\tau_{\tau s}=0.055~\mathrm{s}$ (dashed black), assuming mixing with $\nu_\tau$. Note how the more massive neutrino decays deeper in its non-relativistic regime and well after its decoupling: The initial decline is due to the still-coupled massive neutrino, experiencing `Boltzmann suppression'. Soon later, it decouples and its contribution to the plasma rises, with $N_{\mathrm{eff}}$  that follows this trend. The growth is somewhat less steep than naively expected in absence of decays since this process partially counteracts the sterile neutrino energy density relative growth. This growth turns into a sharp decline around $x\simeq 0.25$, when decay takes over. If all its entropy were transferred to the active neutrinos only, $N_{\mathrm{eff}}$ would stay constant. The partial redistribution to the e.m. plasma causes a minor decline in $N_{\mathrm{eff}}$, basically complete by $x\sim 1$. Modulo quantitative differences, the lighter and longer lived neutrino depicted with the red line follows the same stages but shifted to the right since it stays coupled longer and decays later. 

The contribution of the sterile neutrino to this dynamics is more clearly visible in Fig.~\ref{rhos_100vs30}, while Fig.~\ref{z_100vs30} shows the entropy dilution effect entering $z$, which partially counteracts the extra energy density in neutrinos and also affects $N_{\mathrm{eff}}$. All other parameters being the same, the effect on $z$ is more pronounced, as expected due to the larger branching ratio in e.m. species. 
Note that a further, ``standard'' enhancement in $z$, due to $e^\pm$ annihilation, happens at $x\gg 1$ and is not visible in the figure. In Fig.~\ref{rho_100vs30}  we show the evolution of the active neutrino energy densities. Note how they already depart from equilibrium somewhat by $x\sim 0.2$ (for $m_s=100\,$MeV), when the major enhancement happens due to the bulk of the decays.
%......................................................

%%%%%%%%%%%%%%%%%%%%%%%%%%%%%%%%%%%%%%%%%%
\subsection{Impact on $Y_p$}\label{impactYp}
%%%%%%%%%%%%%%%%%%%%%%%%%%%%%%%%%%%%%%%%%%%
Another important parameter affected by a massive sterile neutrino scenario is the $Y_p$ value, i.e a proxy for the primordial $^4$He mass fraction.  The effect
arises due to the modified expansion history, i.e. via the role that $\neff$ and $z(x)$ play in the Hubble function $H(x)$, but above all via the distortions to the electron (anti)neutrino
distribution entering the isospin changing reactions between neutrons and protons. In Fig.~\ref{distrib}, we illustrate the typical $\nu_\tau$ and $\nu_e$ spectra ($\nu_\mu$ being intermediate between the two) associated to heavy sterile neutrino decays, in the case of mixing with $\nu_\tau$: Despite mixing among the active species, the largest distortion remains in the $\nu_\tau$ species; also, heavier neutrinos lead to more energetic residual distortions. Alterations due to heavy sterile neutrino decays also affect other nuclei such as deuterium, but these are subleading compared to the effect on $Y_p$ (see e.g. Fig. 3 in~\cite{Sabti:2020yrt}) and for simplicity we will limit ourselves to model the modifications on $Y_p$. Both 
CMB and  BBN are sensitive to $Y_p$, but astrophysical determinations of $Y_p$ and thus a comparison with primordial nucleosynthesis predictions is currently more constraining. 

%%%%%%%%%%%%%%%%%%%%%%%%%%%%%%%%%%%%%%%%%%%%%%%%%%%%
A precise standard model calculation, $Y_{p, {\rm SM}}^{\rm prec}$, for the best-fit cosmological parameter  
$\Omega_bh^2=0.02225$~\cite{Aghanim:2018eyx} with a number of subtle effects included (see~\cite{Serpico:2004gx,Iocco:2008va} for details), is available from \texttt{Parthenope}~\cite{Pisanti:2007hk,Consiglio:2017pot}. This result is rescaled via the ratio of the Born estimate of the $Y_p$ for the $\nu_s$ model, $Y_{p, \nu_s}^{\rm Born}$, over the Born standard model calculation $Y_{p, {\rm SM}}^{\rm Born}$, as 
\begin{equation}
Y_p=Y_{p, {\rm SM}}^{\rm prec}\, \frac{Y_{p, \nu_s}^{\rm Born}}{Y_{p, {\rm SM}}^{\rm Born}} \,\ .
\end{equation}
 Each term of the fraction at the r.h.s. can be estimated as (see e.g.~\cite{Esposito:1998rc}):
%%%%%%%%%%%%%%%%%%%%%%%%%%%%%%%%%%%%%
\begin{align}
Y_p=2X_n (t_{\rm on})e^{-t_{\rm on}/\tau_n} \quad\quad\quad X_n=\frac{n_n}{n_p+n_n} \,\ ,\\
\frac{dX_n}{dx}=\frac{\left[\omega_B(p\rightarrow n)\left(1-X_n\right)-\omega_B\left(n\rightarrow p\right)X_n\right]}{xH} \,\ ,
\end{align}
%%%%%%%%%%%%%%%%%%%%%%%%%
where $t_{\rm on}\simeq 180~\mathrm{s}$ corresponds to the onset of the BBN (i.e. deuterium bottleneck opening around $T\simeq 0.08\,$MeV)~\footnote{Note that, while strictly speaking $t_{\rm on}$ is altered in the non-standard scenario considered here, the bulk of the change in $Y_p$ comes from the prefactor $\propto X_n$ in Eq.~(39). A simple estimate yields the scaling $t_{\rm on}\simeq 180\,{\rm s}\sqrt{H_{\rm SM}/H}$, where $H_{\rm SM}$ and $H$ are the Hubble parameter values at the beginning of BBN in the Standard Model and the case under exam, respectively.
For a typical allowed modification of $N_{\mathrm{eff}}=3.2$, the effect on $t_{\rm on}$ is $~2\%$, propagating to a $~0.3\%$ effect on $Y_p$, about one order of magnitude below the 2\,$\sigma$ observational error on $Y_p$ considered in the following, see eq.~\ref{obsYp}. As a consequence, neglecting the change of $t_{\rm on}$ does not lead to appreciably different results.}, $\tau_n$ is the neutron lifetime and $\omega_B$ are the rates in the Born approximation of the processes in Table~\ref{npreactions} (with $\Delta\equiv m_n-m_p\simeq1.29\,$MeV) that can be written as:
%%%%%%%%%%%%%%%%%%%%%%%
\begin{equation}
\omega_B=\frac{G_F^2\left(C_V^2+3C_A^2\right)}{2\pi^3}\int_0^{\infty}dpp^2q_0^2\theta(q_0)F \,\ .
\end{equation}
The distributions entering in $F$, as well as $H(x)$, are taken from the numerical solutions of our system of equations.
In the last column of Tab.~\ref{TD}, we show the results for some relevant sterile neutrino parameters.
%%%%%%%%%%%%%%%%%%%%%%%%%%%%%%
\begin{table}
\centering
\caption{Relevant quantities for $n-p$ reactions}
\label{npreactions}
\begin{tabular}{lcr}
\hline
Process  &$F$ &$q_0$\\
\hline
\hline
$\nu_e+n\rightarrow e^-+p$ &$f_{\nu}(q_0)\left(1-f_e(p_0)\right)$ &$-\Delta+p_0$ \\
$e^-+p\rightarrow\nu_e+n$ &$f_e(p_0)\left(1-f_{\nu}(q_0)\right)$ &$-\Delta+p_0$\\
$e^++p\rightarrow\bar{\nu}_e+p$ &$f_e(p_0)\left(1-f_{\nu}(q_0)\right)$ &$\Delta+p_0$\\
$\bar{\nu}_e+p\rightarrow e^++n$ &$f_{\nu}(q_0)\left(1-f_e(p_0)\right)$ &$\Delta+p_0$ \\
$n\rightarrow e^-+\bar{\nu}_e+p$ &$\left(1-f_{\nu}(q_0)\right)\left(1-f_e(p_0)\right)$ &$\Delta-p_0$\\
$e^-+\bar{\nu}_e+p\rightarrow n$ &$f_{\nu}(q_0)f_e(p_0)$ &$\Delta-p_0$\\
\hline
\end{tabular}
\end{table}

%%%%%%%%%%%%%%
%%%%%%%%%%%%%%%%%%%%%

%...........................
%\textcolor{red}{
%\begin{table}
%\caption{Results for the $Y_p$ value. $\tau$ is the lifetime of the sterile neutrino considered.}
%\vspace{.5cm}
%\centering
%\begin{tabular}{cccc}
%\hline
%$m_s~[\mathrm{MeV}]$&  $\sin^2\theta_{\tau 4}$& $\tau~[\mathrm{s}]$& $Y_p$\\
%\hline
%$20.0$& $2.6\times10^{-2}$& $3.0\times 10^{-1}$& $0.2514$\\
%$40.0$& $2.8\times10^{-3}$& $8.8\times 10^{-2}$& $0.2520$\\
%$60.0$& $5.5\times10^{-4}$& $6.0\times 10^{-2}$& $0.2509$\\
%$80.0$& $1.5\times10^{-4}$& $5.0\times 10^{-2}$& $2---$\\
%$100.0$& $5.8\times10^{-5}$& $4.4\times 10^{-2}$& $---$\\
%$130.0$& $1.6\times10^{-5}$& $4.2\times 10^{-2}$& $---$\\
%\hline
%\end{tabular}
%\label{Yptable}
%\end{table}
%..................................
%}

%%%%%%%%%%%%%%%%%%%%%%%%%%
\section{Constraints and forecasts}\label{C&F}
%%%%%%%%%%%%%%%%

In order to obtain constraints on heavy sterile neutrinos, we compare our results on $N_{\rm eff}$ and
on the modification on $Y_p$ with both the latest CMB and BBN measurements. For BBN, we use the current bound at $2\sigma$~\cite{Tanabashi:2018oca}
%%%%%%%%%%%%%%%%%%%%%%%%%%%%%%%%%%%%%%%

\begin{equation}
Y_p=0.245\pm 0.006 \,\ .\label{obsYp}
\end{equation}

Concerning CMB, if limiting oneself to $N_{\rm eff}$, the latest measurements of the Planck collaboration provide a value 
$N_{\rm eff}=2.99 \pm 0.17$~\cite{Aghanim:2018eyx}. Therefore we could exclude at $2\sigma$ 
extra-radiation leading to
$\Delta N_{\rm eff}> 0.33$.
In practice, the massive sterile neutrino model under consideration here leads to changes in both $N_{\rm eff}$ and $Y_p$, and the CMB is sensitive to both (albeit much less to $Y_p$ than BBN, at the moment).
Hence we infer the CMB constraints using a reduced Gaussian likelihood matrix 
involving $N_{\mathrm{eff}}$ and $Y_p$, of the form~\cite{Sabti:2020yrt}:

%%%%%%%%%
\begin{figure}
\centering
\includegraphics[scale=0.6]{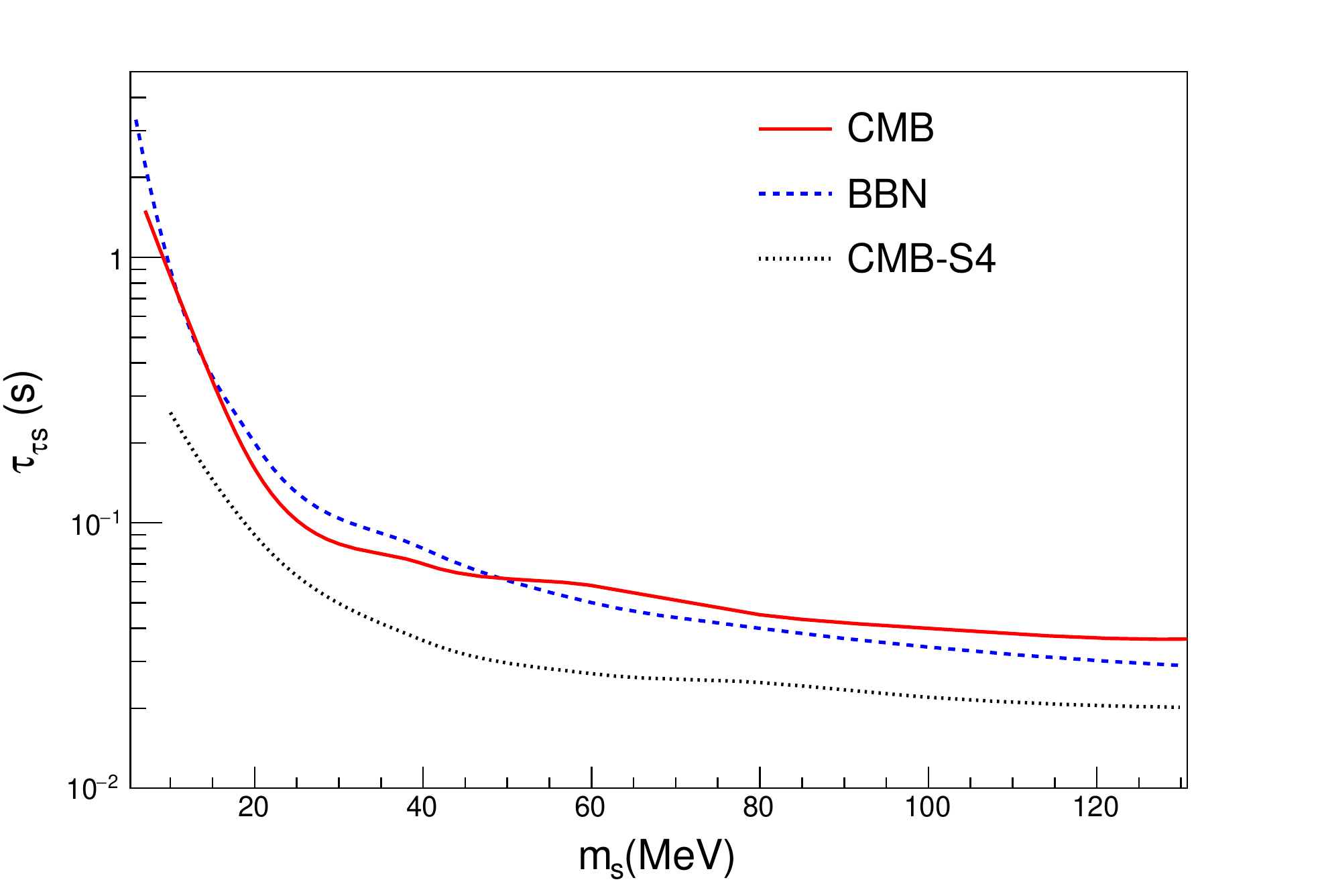}
\caption{Bounds in the plane $(m_s,\tau_s)$ obtained from CMB (red curve) and BBN-$Y_p$ (blue curve), as well as forecast sensitivity of CMB-S4 (black curve), for a sterile neutrino mixed with  $\nu_\tau$ (or $\nu_\mu$). The $2\sigma$ excluded region is the one above the curves.}
\label{bound_t}
\end{figure}
%%%%%%%

%%%%%%%%%
\begin{figure}
\centering
\includegraphics[scale=0.6]{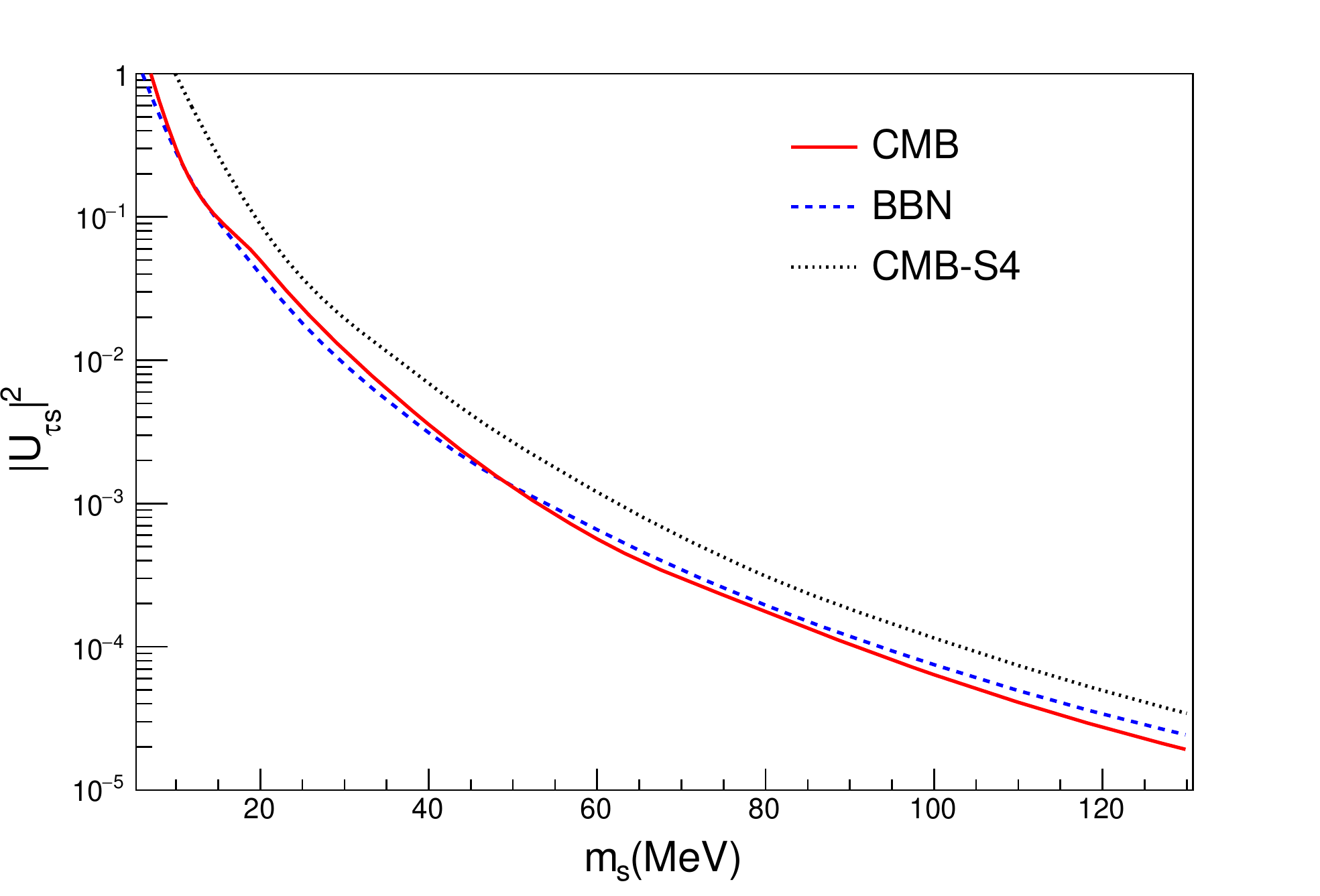}
\caption{Bounds in the plane $(m_s,\theta_{\tau s})$ obtained from CMB (red curve) and BBN-$Y_p$ (blue curve), as well as forecast sensitivity of CMB-S4 (black curve), for a sterile neutrino mixed with  $\nu_\tau$ (or $\nu_\mu$). The $2\sigma$ excluded region is the one under the curves.}
\label{bound_t_theta}
\end{figure}
%%%%%%%

%%%%%%%%%
\begin{figure}
\centering
\includegraphics[scale=0.6]{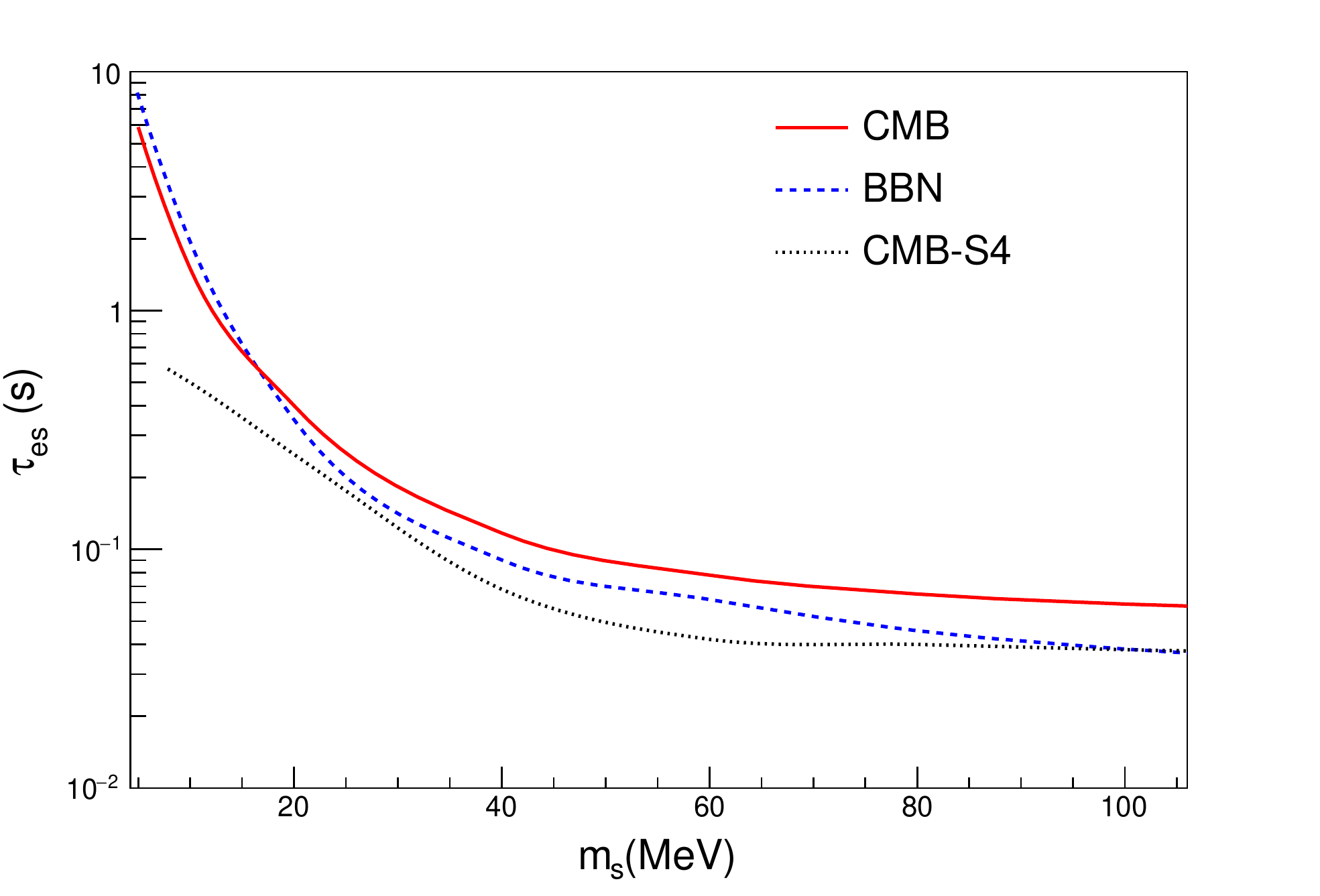}
\caption{Bounds in the plane $(m_s,\tau_s)$ obtained from CMB (red curve) and BBN-$Y_p$ (blue curve), as well as forecast sensitivity of CMB-S4 (black curve), for a sterile neutrino mixed with  $\nu_e$. The $2\sigma$ excluded region is the one above the curves.}
\label{bound_e}
\end{figure}
%%%%%%%

%%%%%%%%%
\begin{figure}
\centering
\includegraphics[scale=0.6]{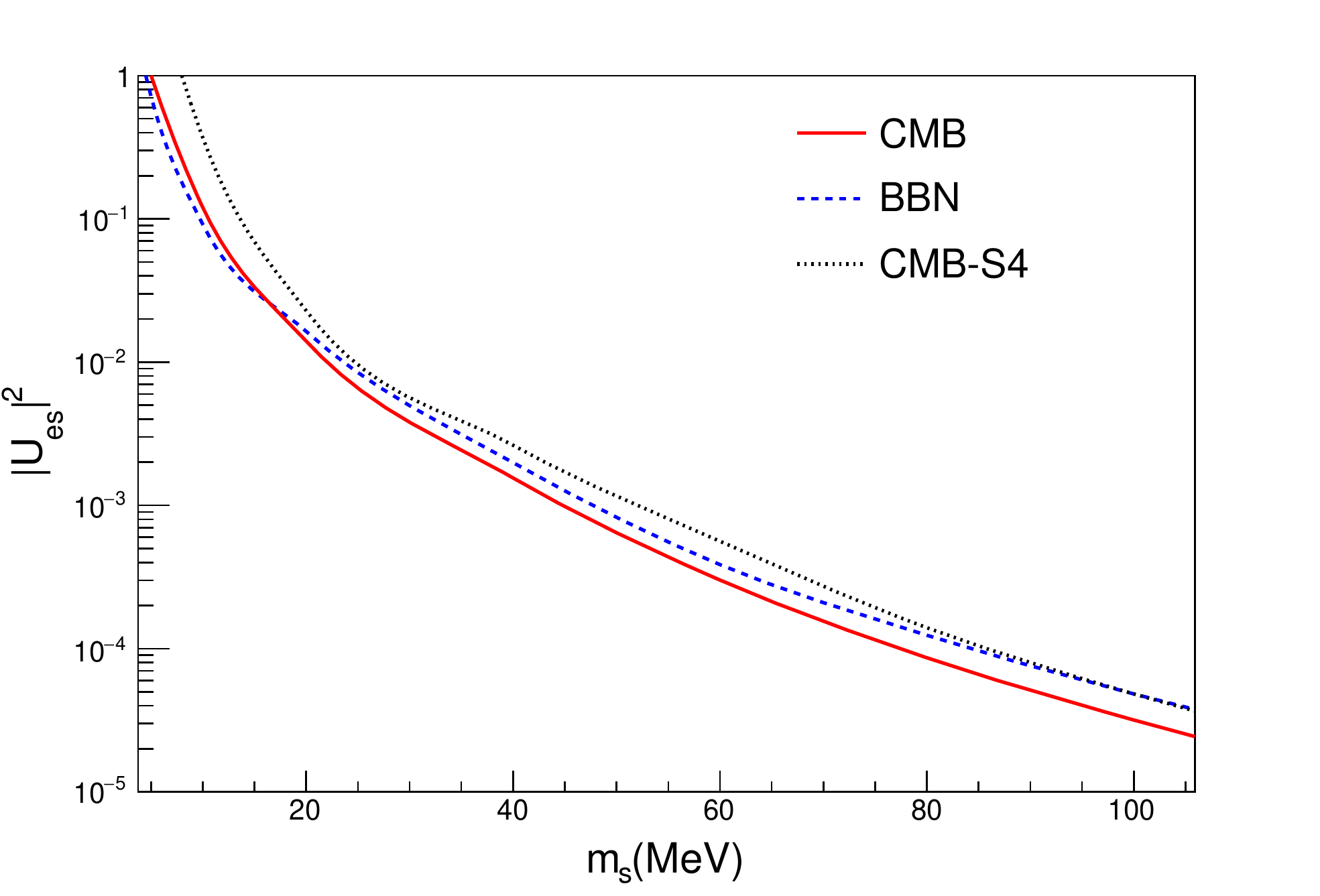}
\caption{Bounds in the plane $(m_s,\theta_{es})$ obtained from CMB (red curve) and BBN-$Y_p$ (blue curve), as well as forecast sensitivity of CMB-S4 (black curve), for a sterile neutrino mixed with  $\nu_e$. The $2\sigma$ excluded region is the one under the curves.}
\label{bound_e_theta}
\end{figure}
%%%%%%%

%%%%%%%%%%%%%%%%%%%%%%%%%%%%%%%%%%%%%%%%%%%%%%%%%%%%%%%%%%%%%%%%
\begin{align}
\chi^2_{\mathrm{CMB}}&=\left(\Theta-\Theta_{\mathrm{obs}}\right)\Sigma^{-1}_{\mathrm{CMB}}\left(\Theta-\Theta_{\mathrm{obs}}\right)^T \,\ ,\\
\Theta&=\left(N_{\mathrm{eff}},Y_p\right) \,\ , \\
\Theta_{\mathrm{obs}}&=\left(2.97,0.246\right) \,\ , \label{osservato}\\
\Sigma_{\mathrm{CMB}}&=\left(
\begin{matrix}
 \sigma_1^2& &\sigma_1\sigma_2\rho_{12} \\
 \sigma_1\sigma_2\rho_{12}& &\sigma_2^2
\end{matrix}
\right) \,\ , \\
\left(\sigma_1,\sigma_2\right)&=(0.2650,0.0177) \,\ , \\
\rho_{12}&=-0.845 \,\ .
\end{align}
%%%%%%%%%%%%%%%%%%%%%%%%%%%%%%%%%%%%%%%%%%%%%%%%%%%%%%%%%%%%%%%%
To obtain results at $2\sigma$, we have to consider a value of $\chi^2=6.18$, value obtained by requiring that the integral of the $\chi^2$-distribution with 2 dof is equal to $0.9545$. Our results from CMB measurements and from BBN  based on $Y_p$ are shown in Fig.~\ref{bound_t} for the constraints on the decay time, and in  Fig.~\ref{bound_t_theta}  for the constraints on the mixing parameter, for the most interesting case of mixing with $\nu_\tau$. 
As a side result and in order to allow for cross-checks with past literature, in Fig.~\ref{bound_e} and Fig.~\ref{bound_e_theta} we also report the corresponding results for a mixing with $\nu_e$.
Besides the current constraints, we also show the sensitivity forecast of the future CMB-S4 observations, with uncertainties $\left(\sigma_1,\sigma_2\right)=(0.062,0.0053)$ according to~\cite{Baumann:2015rya}, considering the same $\Theta_{\mathrm{obs}}$ as in Eq.~(\ref{osservato}).

We conclude that the CMB provides already the best constraints for  $m_s\lesssim 50\,$MeV, while BBN takes over at larger masses. However, we expect that CMB-S4 will attain leading constraining power in the whole range of parameter space considered here, if performing close to expectations. 
Qualitatively similar consideration apply for mixing with $\nu_e$, although the transition mass is around  $m_s\sim 20\,$MeV, and the future improvement of CMB-S4 over BBN is less significant. Note that, in particular for the CMB, at the same mass and lifetime the bounds are more stringent for a mixing with $\nu_\tau$ than one with $\nu_e$: This is due to the fact that the bound is dominated by $N_{\rm eff}$ and, due to the larger b.r. in neutrinos for the case of mixing with $\nu_\tau$, the growth of neutrino density via non-thermal injection is only mildly compensated by the entropy effect. For the case of mixing with  $\nu_e$, there is instead a substantial compensation via the growth of $z$. 

In the case of the BBN bound, however, the leading effect is due to $\nu_e$ distortions, which are larger when the mixing is with $\nu_e$; the effect of $N_{\rm eff}>3$ altering $H$ is however more relevant when the mixing is with $\nu_\tau$, so that the two constraints are closer to each other in this case. 

Our BBN constraints are largely consistent with recent calculations presented in~\cite{Sabti:2020yrt}, while our CMB constraints are consistent with theirs at low masses, while more stringent than theirs at high masses. A more detailed comparison and discussion are reported in Appendix~\ref{appC}, where we identify the origin of the difference in the estimate of $\Delta N_{\mathrm{eff}}$. In particular, contrarily to the results of~\cite{Sabti:2020yrt}, we always obtain $\Delta N_{\mathrm{eff}}>0$ for the parameter space of interest. We have supplemented our numerical calculations with a qualitative study of the Boltzmann equation in the analytical approximation of~\cite{Dolgov:2000jw} to further support our conclusions.
%%%%%%%%%%%%%%%%%%%%%%%%%%%%%%%%%%%%%%%
\section{Conclusions}\label{summary}

Heavy sterile neutrinos with masses $\mathcal{O}$(MeV-GeV) are predicted in extensions of the Standard Model such as the Neutrino Minimal Standard Model  ($\nu$MSM). Besides affecting collider and supernovae observables, their mass and mixing angle parameters can be also constrained with cosmological observables, notably  CMB and BBN.

We have numerically studied the evolution of sterile neutrinos with $10~\mathrm{MeV}\lesssim m_s \lesssim 135~\mathrm{MeV}$ in the early universe and set constraints on the mixing angles or lifetimes using the $N_{\mathrm{eff}}$ and $Y_p$ observables. In order to achieve these results, we have solved the exact Boltzmann equation for sterile and active neutrino evolution while taking into account the temperature evolution of electrons and photons. Also, we checked the correctness of  analytical approximations in the literature and verified that they are adequate to describe sterile neutrino decoupling at better than 10\% level. 

For the least constrained (and thus phenomenologically most interesting) sector of mixing with $\nu_\tau$, at $m_s\gtrsim 50\,$MeV  these cosmological bounds surpass the traditional benchmark of 0.1 s lifetime often considered in the literature, up to about $0.03~\mathrm{s}$ for the highest masses considered. While currently CMB is more constraining at low masses and BBN dominates at high masses, we expect the future CMB-S4 experiments to yield the dominant constraining power by the end of the decade, unless the systematic error affecting the astrophysical determinations of $Y_p$ can be significantly reduced.
%%%%%%%%%%%%%%%%%%%%%%%%%%%%%%%%%%%%%%%%
\section*{Acknowledgements}
%%%%%%%%%%%%%%%%%%%%%%%
We warmly thank Gennaro Miele for discussions during the development of this project, as well as A. Boyarsky, P. Fernandez de Salas, S. Hannestad, S. Pastor, and A. Sabti for fruitful exchanges. The work of A.M. and N.S.  is partially supported by the Italian Istituto Nazionale di Fisica Nucleare (INFN) through the ``Theoretical Astroparticle Physics'' project and by the research grant number 2017W4HA7S ``NAT-NET: Neutrino and Astroparticle Theory Network'' under the program
PRIN 2017 funded by the Italian Ministero dell'Universit\`a e della
Ricerca (MUR).\\
The work of L.M. is supported by the Italian Istituto Nazionale di Fisica Nucleare (INFN) through the ``QGSKY'' project and by Ministero dell'Istruzione, Universit\`a e Ricerca (MIUR).\\
The computational work has been executed on the IT resources of the ReCaS-Bari data center, which have been made available by two projects financed by the MIUR (Italian Ministry for Education, University and Research) in the "PON Ricerca e Competitività 2007-2013" Program: ReCaS (Azione I - Interventi di rafforzamento strutturale, PONa3\_00052, Avviso 254/Ric) and PRISMA (Asse II - Sostegno all'innovazione, PON04a2A).

\appendix

%%%%%%%%%%%
\section{Reduction of the decay and collisional integrals}\label{appA}
%%%%%%%%%%%
In this Appendix, we show how, under the approximations adopted in~\cite{Dolgov:2000jw}, we can  analytically solve the integrals appearing at
the r.h.s. of Eq.~(\ref{eq:sterile}), namely
%%%%%%%%%%%%%%
\begin{eqnarray}
I_{\mathrm{coll}}+I_{\mathrm{dec}}&=&\frac{1}{2E}\int\prod_i\left(\frac{d^3p_i}{2E_i(2\pi)^3}\right)\prod_f\left(\frac{d^3p_f}{2E_f(2\pi)^3}\right)\nonumber \\
&\times& (2\pi)^4\delta^{(4)}\left(\sum_i p_i-\sum_fp_f\right)|M_{fi}|^2F(f_i,f_f) \,\ , 
\label{general_coll_integral}
\end{eqnarray}
%%%%%%%%%%%%%%%%%%%%%%%

with $|M_{fi}|^2$ the  sum of the squared-matrix elements for  the decay and scattering processes
 and
%%%%%%%%%%%%%
\begin{equation}
F(f_i,f_f)=-\prod_if_i\prod_f(1-f_f)+\prod_i(1-f_i)\prod_ff_f \,\ .
\label{eq:kernel}
\end{equation}
%%%%%%%%%%%%%%%%%%%%%%
Please note that our results are obtained with a numerical integration of the collisional integrals in Eq.~(\ref{general_coll_integral}), after the dimensional reduction recapped in Appendix~\ref{appB}, not with the analytical approximations.
Also, None of the approximations reported below is new and can be skipped unless one is interested in reproducing the analytical approximations. However, for pedagogical purposes, we thought it is useful to report them here in great detail, not to force readers to go back to the decades-old original literature. 

%%%%%%%%%%%%%%%%%
\subsection{Decay Integral}
%%%%%%%%%%%%%%%%%

Following the approximations of~\cite{Dolgov:2000jw} we assume that in Eq.~(\ref{general_coll_integral})
the energy distributions are represented by Maxwell-Boltzmann distributions, instead of Fermi-Dirac ones. 
Moreover, we neglect the Pauli blocking factor, assuming  $(1-f_i)\simeq1$. Then one gets
%%%%%%%%%%%%%%%%
\begin{equation}
I_{\mathrm{dec}}=\frac{1}{2E_1(2\pi)^5}\int\frac{{\rm d}^3p_2}{2E_2}\frac{{\rm d}^3p_3}{2E_3}\frac{{\rm d}^3p_4}{2E_4}|M_{fi}|^2[-f_1+e^{-E_2/T}e^{-E_3/T}e^{-E_4/T}]\delta^{(4)}(p_1-p_2-p_3-p_4) \,\ ,
\label{eq:dec}
 \end{equation} 
 %%%%%%%%%%%%%%%%%
 where the label $1$ indicated the sterile neutrino and
 $|M_{fi}|^2$ is the sum over the dominant decay processes
 %%%%%%%%%%%
\begin{align*}
\nu_s&\rightarrow \nu_{\tau}\nu_{\alpha}\bar{\nu}_\alpha \,\ , \\
\nu_s&\rightarrow\nu_{\tau}e^+e^- \,\ ,
\end{align*} 
%%%%%%%%%%%%%
with $\alpha=e,\mu,\tau$. Performing the integral over d$^3p_4$ in eq. (\ref{eq:dec}) using the delta function enforcing $E_1=E_2+E_3+E_4$, one obtains that $e^{-E_2/T}e^{-E_3/T}e^{-E_4/T}=e^{-E_1/T}=f_1^{\rm eq}$. Moreover, using the property of the delta function in eq.~(\ref{deltaprop}), we can write
%%%%%%%%%%
 \begin{equation}
  I_{\mathrm{dec}}=\left(f_1^{\mathrm{eq}}-f_1\right)\frac{1}{2E_1(2\pi)^5}\int\frac{{\rm d}^3p_2}{2E_2}\frac{{\rm d}^3p_3}{2E_3}|M_{fi}|^2\delta((p_1-p_2-p_3)^2)= \left(f_1^{\mathrm{eq}}-f_1\right)\frac{m_1}{E_1}\Gamma_D \,\ ,
  \end{equation}
  %%%%%%%%%%%%%%%%%
  where the sterile neutrino decay rate $\Gamma_D$ is given by~\cite{Dolgov:2000pj}
  %%%%%%%%%%%%
  \begin{equation}
  \begin{split}
  \Gamma_D&\equiv\frac{1}{2m_1(2\pi)^5}\int\frac{d^3p_2}{2E_2}\frac{{\rm d}^3p_3}{2E_3}\frac{{\rm d}^3p_4}{2E_4}|M_{fi}|^2\delta(p_1-p_2-p_3-p_4)\\&=\frac{1}{2m_1(2\pi)^5}\int\frac{{\rm d}^3p_2}{2E_2}\frac{{\rm d}^3p_3}{2E_3}|M_{fi}|^2\delta((p_1-p_2-p_3)^2)=\frac{(1+\tilde{g}_L^2+g_R^2)G_F^2m_1^5|U_{s\tau}|^2}{192\pi^3} \,\ .
  \end{split}
  \end{equation}
  %%%%%%%%%%%%%%%%%%
  Thus, we end up with 
%%%%%%%%%%%%%%
\begin{equation}
I_{\mathrm{dec}}=\frac{m_s}{E_s}\frac{1}{\tau_s}(f_s^{eq}-f_s) \,\ ,
\end{equation}
%%%%%%%%%%%%%%%%%%%%%%%%%%%
where $\tau_s=1/\Gamma_D$ is the sterile neutrino lifetime.

We could also write the corresponding source term for an active species at the same level of approximation (i.e. neglecting the blocking factor and the inverse decay) following the treatment detailed in Ref.~\cite{Mastrototaro:2019vug}, which we address the reader to for details. This yields 
\begin{equation}
H\,x\,\frac{df_a}{dx}(x,y)\simeq\sum_i\frac{B_i}{\tau_s}\int {\rm d\cos\theta}\int_{0}^{\infty}{\rm d}y_s\frac{y_s^2}{y^2}\left(f_s(x,y_s)-f_s^{\mathrm{eq}}(x,y_s)\right)\mathcal{F}_{a,i}\left(\frac{y}{\gamma(1+\beta\cos\theta)},\cos\theta\right) \,\ ,\label{apprsouterm}
\end{equation} 
where $B_i$ is the branching ratio of the i-th exclusive reaction, $\gamma=\sqrt{1+(y_sm)^2/(x m_s)^2}$, $\beta=y_s/\sqrt{(m_sx/m)^2+y^2_s}$, and $\mathcal{F}_{a,i}$ is the double-differential distribution (with respect to $y$ and to the angular variable $\theta$) of the daughter particle $a$ in the reaction $i$ in the sterile neutrino rest frame, normalized to 1. The integral kernel in the integral above acts as a constraint picking the ``right'' momentum for the daughter neutrino, weighting it for the occupation factor of the parent sterile species. In practice, we always use the full numerical integration rather than this approximate expression. 
%%%%%%%%%%%%%%%%%%%%%%%%%%%%%%%%%
\subsection{Collisional Integral}
%%%%%%%%%%%%%%%%%%%%%%%%%%%%%%%%%

In order to evaluate the collisional integral in Eq.~(\ref{general_coll_integral}), we 
assume below that they only mix with $\nu_\tau$. 
The relevant collisional processes are shown in Table \ref{scattering} and have a squared interaction matrix given by
%.................
\begin{equation}
 |M|^2 =  4C[(p_1\cdot p_2)(p_3\cdot p_4)+2(p_1\cdot p_4)(p_2\cdot p_3)] \,\ ,
\end{equation}
%.........................
where $C=2^{4}G_F^{2}|U_{\tau 4}|^{2}(1+\tilde{g}_L^2+g_R^2)$ (remember that  in our definition of $|M|^2$ we sum over all the degrees of freedom and include the average over the
relevant state). We indicate with $1$ the sterile neutrino state, and evaluate the quantities in the center of momentum frame; since particles 2, 3, 4 are  relativistic, 
${\bf p}_1=-{\bf p}_2$, ${\bf p}_3=-{\bf p}_4$, $E_3=E_4$, $E_1+E_2=E_3+E_4=2E_3=2E_4$. We have
\begin{equation}
p_1\cdot p_2=E_1E_2-{\bf p}_1\cdot{\bf p}_2=E_1E_2+{\bf p}_1^2=E_1E_2+\frac{{\bf p}_1^2+{\bf p}_2^2}{2}=E_2(E_1+{\bf p}_1)
\end{equation}
and
\begin{equation}
p_3\cdot p_4=E_3E_4-{\bf p}_3\cdot{\bf p}_4=2E_3E_4=2E_3^2.
\end{equation}
Thus
\begin{equation}
p_1\cdot p_2-p_3\cdot p_4=E_1E_2+\frac{{\bf p}_1^2+{\bf p}_2^2}{2}-2\frac{(E_1+E_2)^2}{4}=
\frac{{\bf p}_1^2-E_1^2}{2}+\frac{{\bf p}_2^2-E_2^2}{2}=-\frac{m_s^2}{2}.
\end{equation}
Hence:
\begin{equation}
(p_1\cdot p_2)(p_3\cdot p_4)=  (p_1\cdot p_2)^2 +\frac{m_s^2}{2}(p_1\cdot p_2)=(p_1\cdot p_2)^2 +\frac{m_s^2}{2}[(p_3\cdot p_2)+(p_4\cdot p_2)]\,.
\end{equation}
Similarly, 
\begin{eqnarray}
p_1\cdot p_4-p_2\cdot p_3&=&E_1E_4-E_2E_3-{\bf p}_1\cdot{\bf p}_4+{\bf p}_2\cdot{\bf p}_3\nonumber\\
&=&E_4(E_1-E_2)+{\bf p}_3\cdot({\bf p}_1+{\bf p}_2)\nonumber\\
&=&\frac{(E_1+E_2)}{2}(E_1-E_2)+{\bf 0}=\frac{(E_1^2-E_2^2)}{2}=\frac{m_s^2}{2}\,.
\end{eqnarray}
Hence:
\begin{equation}
(p_1\cdot p_4)(p_2\cdot p_3)=  (p_2\cdot p_3)^2 +\frac{m_s^2}{2}(p_2\cdot p_3)\,,
\end{equation}
These equalities among Lorentz invariants hold in any frame. As a result, let us write:
\begin{equation}
 |M|^2 = 4\,C\left[ (p_1\cdot p_2)^2+2 (p_2\cdot p_3)^2  +\frac{m_s^2}{2}[3(p_3\cdot p_2)+(p_4\cdot p_2)] \right]\equiv I[{\sf A}]+I[{\sf B}] + I[{\sf C}] \,\ .
\end{equation}
We thus have
\begin{equation}
\begin{split}
\frac{I[{\sf A}]}{4\,C}=&\int\frac{{\rm d}^3p_2{\rm d}^3p_3{\rm d}^3p_4}{(2\pi)^92^4E_1E_2E_3E_4}(p_1\cdot p_2)^2\delta(p_1+p_2-p_3-p_4)(2\pi)^4\left(f^{\mathrm{eq}}(E_1)-f(E_1)\right)f^{\mathrm{eq}}(E_2)\\=&
\int\frac{{\rm d}^3p_2{\rm d}^3p_3}{(2\pi)^52^3E_1E_2E_3}(p_1\cdot p_2)^2\delta((p_1+p_2-p_3)^2)\left(f^{\mathrm{eq}}(E_1)-f(E_1)\right)f^{\mathrm{eq}}(E_2)\\=&
\int\frac{{\rm d}^3p_2{\rm d}^3p_3}{(2\pi)^52^3E_1E_2E_3}(p_1\cdot p_2)^2\delta({m_s^2+}2p_1\cdot p_2-2p_2\cdot p_3-2p_3\cdot p_1)\left(f^{\mathrm{eq}}(E_1)-f(E_1)\right)f^{\mathrm{eq}}(E_2)\\=&
\int\frac{{\rm d}^3p_2{\rm d}^3p_3}{(2\pi)^52^4E_1E_2E_3}(p_1\cdot p_2)^2\left(f^{\mathrm{eq}}(E_1)-f(E_1)\right)f^{\mathrm{eq}}(E_2)\\&\delta\left({\frac{m_s^2}{2}+}(p_1\cdot p_2)-E_2E_3(1-\cos\theta_{32})-E_1E_3(1-{v_1}\cos\theta_{31})\right)\\=&
\int\frac{{\rm d}^3p_2{\rm d}E_3{\rm d}\cos\theta_{31}{\rm d}\mu_3E_3}{(2\pi)^52^4E_1E_2}(p_1\cdot p_2)^2\frac{1}{E_2(1+\cos\theta_{31})+E_1(1-{v_1}\cos\theta_{31})}\\
 &\delta\left(E_3-\frac{{(p_1\cdot p_2)+m_s^2/2}}{E_2(1+\cos\theta_{31})+E_1(1-{v_1}\cos\theta_{31})}\right)\left(f^{\mathrm{eq}}(E_1)-f(E_1)\right)f^{\mathrm{eq}}(E_2)\\=&
\int\frac{{\rm d}^3p_2}{(2\pi)^42^4E_1E_2}\left[(p_1\cdot p_2)^3+(p_1\cdot p_2)^2\frac{m_s^2}{2}\right]\left(f^{\mathrm{eq}}(E_1)-f(E_1)\right)f^{\mathrm{eq}}(E_2)\\
&\times\int\frac{{\rm d}\mu_3}{2\pi}\int {\rm d}\cos\theta_{31}\frac{1}{\left[E_2(1+\cos\theta_{31})+E_1(1-{v_1}\cos\theta_{31})\right]^2}\\=&
\int\frac{{\rm d}^3p_2}{(2\pi)^42^4E_1E_2}\frac{2(p_1\cdot p_2)^3+(p_1\cdot p_2)^2{m_s^2}}{2(p_1\cdot p_2){+m_s^2}}\left(f^{\mathrm{eq}}(E_1)-f(E_1)\right)f^{\mathrm{eq}}(E_2)\\
=&\int\frac{{\rm d}^3p_2}{(2\pi)^42^4E_1E_2}(p_1\cdot p_2)^2\left(f^{\mathrm{eq}}(E_1)-f(E_1)\right)f^{\mathrm{eq}}(E_2)\,\ ,
\end{split}
\label{E_A}
\end{equation}
where $\theta$ and $\mu$ are the azimuth and polar angle respectively,  $\cos\theta_{32}=-\cos\theta_{31}$ due to the fact we evaluated the integral in the center of momentum frame, and we used Eq.~(\ref{int1b}), with $a=E_1+E_2$ and $b=E_2-|{\bf p}_1|$, as well as
\begin{equation}
\delta(\kappa x)=\frac{1}{|\kappa|}\delta(x) \,\ .
\end{equation}
Let us evaluate  the integral over particle 2 (expressed in terms of Lorentz-invariants) by the explicit replacement $p_1\cdot p_2=E_1E_2-p_1p_2\cos\theta_{12}$:
\begin{equation}
\begin{split}
\frac{I[{\sf A}]}{4\,C}=&\left(f^{\mathrm{eq}}(E_1)-f(E_1)\right)\int\frac{{\rm d}E_2{\rm d}\cos\theta_{12}{\rm d}\mu_2}{(2\pi)^42^4E_1}E^3_2(E_1-|{\bf p}_1|\cos\theta_{12})^2f^{\mathrm{eq}}(E_2)\\=&\frac{\left(f^{\mathrm{eq}}(E_1)-f(E_1)\right)}{(2\pi)^32^4E_1}2\left(E_1^2+\frac{|{\bf p}_1|^2}{3}\right)\int {\rm d}E_2E_2^3f^{\mathrm{eq}}(E_2) \,\ .
\end{split}
\end{equation}
Similarly
\begin{equation}
\begin{split}
\frac{I[{\sf B}]}{8\,C}=
&\int\frac{{\rm d}^3p_2{\rm d}^3p_3{\rm d}^3p_4}{(2\pi)^92^4E_1E_2E_3E_4}(p_2\cdot p_3)^2\delta(p_1+p_2-p_3-p_4)(2\pi)^4\left(f^{\mathrm{eq}}(E_1)-f(E_1)\right)f^{\mathrm{eq}}(E_2)\\=
&\int\frac{{\rm d}^3p_2}{(2\pi)^42^4E_1E_2}\frac{(p_1\cdot p_2)^2}{3}\left(f^{\mathrm{eq}}(E_1)-f(E_1)\right)f^{\mathrm{eq}}(E_2)=\frac{1}{3}\frac{I[{\sf A}]}{4\,C}\,\Rightarrow I[{\sf B}]=\frac{2}{3}I[{\sf A}]
\end{split}
\label{E_B}
\end{equation}
%%%%%%%%%%%%%%%%%%%%%%%%%%%%%%%%%%%%%%%%%%%%%%%%%%%
Finally, taking into account that the integrals for the ${\sf C}-$term are symmetric under the relabelling $3\leftrightarrow 4$, we have with a similar procedure of Eq.~(\ref{E_A})
%%%%%%%%%%%%%%%%%%%%%%%%%%%%%%%%%%%%%%%%%%%
\begin{equation}
\begin{split}
\frac{I[{\sf C}]}{8\,C\,m_s^2}=
&\int\frac{{\rm d}^3p_2{\rm d}^3p_3{\rm d}^3p_4}{(2\pi)^92^4E_1E_2E_3E_4}(p_2\cdot p_3)\delta(p_1+p_2-p_3-p_4)(2\pi)^4\left(f^{\mathrm{eq}}(E_1)-f(E_1)\right)f^{\mathrm{eq}}(E_2)\\=
&\int\frac{{\rm d}^3p_2}{(2\pi)^42^4E_1E_2}\frac{(p_1\cdot p_2)}{2}\left(f^{\mathrm{eq}}(E_1)-f(E_1)\right)f^{\mathrm{eq}}(E_2)\,.
\end{split}
\label{E_C}
\end{equation}

Hence
\begin{equation}
\begin{split}
\frac{I[{\sf C}]}{8\,C\,m_s^2}=&\left(f^{\mathrm{eq}}(E_1)-f(E_1)\right)\int\frac{{\rm d}E_2{\rm d}\cos\theta_{12}{\rm d}\mu_2}{(2\pi)^42^4}\frac{E_2^2}{2}\left(1-v_1\cos\theta_{12}\right)f^{\mathrm{eq}}(E_2)\\=&\frac{\left(f^{\mathrm{eq}}(E_1)-f(E_1)\right)}{(2\pi)^32^4}\int {\rm d}E_2E_2^2f^{\mathrm{eq}}(E_2) \,\ .
\end{split}
\end{equation}
Summing all contributions, we have:
\begin{equation}
\frac{I_{\rm coll}}{C} = \frac{\left(f^{\mathrm{eq}}(E_1)-f(E_1)\right)}{(2\pi)^3}\left[\frac{5}{6}\left(E_1+\frac{|{\bf p}_1|^2}{3\,E_1}\right)\int {\rm d}E_2E_2^3f^{\mathrm{eq}}(E_2)+\frac{m_s^2}{2}\int {\rm d}E_2E_2^2f^{\mathrm{eq}}(E_2)\right]\, .
\end{equation}
Assuming in this last step Fermi-Dirac distribution for particle 2, following the procedure in Ref~\cite{Dolgov:2000jw}, 
\begin{equation}
\frac{I_{\rm coll}}{C} = \frac{\left(f^{\mathrm{eq}}(E_1)-f(E_1)\right)}{(2\pi)^3}\left[\left(E_1+\frac{|{\bf p}_1|^2}{3\,E_1}\right)\frac{7\pi^4}{144}T^4+m_s^2\frac{3\zeta(3)}{4}T^3\right]\,,
\end{equation}
which implies
\begin{equation}
I_{\rm coll}=G_F^{2}|U_{\tau 4}|^{2}(1+\tilde{g}_L^2+g_R^2)\left(f^{\mathrm{eq}}(E_1)-f(E_1)\right)T^3\left[\left(E_1+\frac{|{\bf p}_1|^2}{3\,E_1}\right)\frac{7\pi^4}{72}T+\frac{3\zeta(3)}{2\pi^3}m_s^2\right]\,.
\end{equation}
The above result agrees with what reported in ref.~\cite{Dolgov:2000jw} in the same limit:
\begin{eqnarray}
I_{\mathrm{coll}}&=&\frac{4 G_F^2|U_{s\tau}|^2(1+\tilde{g}_L^2+g_R^2)T^3m_s^2}{\pi^3}\left(f^{\mathrm{eq}}(E_1)-f(E_1)\right)\left[ \frac{3}{4}\zeta(3)+\frac{7T\pi^4}{144}\left(\frac{E_1}{m_s^2}+\frac{p_1^2}{3E_1m_s^2}\right)\right] \,\ \nonumber \\
&=&\frac{3\times 2^6}{\tau_s}\left(f^{\mathrm{eq}}(E_1)-f(E_1)\right)\left[3\zeta(3) \frac{T^3}{m_s^3}+\frac{7\pi^4}{36}\frac{T^4E_1}{m_s^5}\left(1+\frac{p_1^2}{3E_1^2}\right)\right],
\end{eqnarray}
%%%%%%%%%%%%%%%%%%%%%%%%%%%%%%%%%%%%%%%%%%%%%%%%%%%%%%%%%%%%%%%%%%%%%%%%%%%%%%%%%
with $\tau_s$ given in Eq.~(\ref{eq:decay}).\\

Finally, we report below some relations used in the numerous integrations:
\begin{equation}
\int \frac{{\rm d }^3 p }{2\,E} \ldots =\int \frac{{\rm d }^3 p{\rm d }p_0 }{2E} \delta(p^0-E)=\int {\rm d }^4 p \, \delta(p\cdot p-m^2)\Theta(p^0)\ldots 
\end{equation}
so that, when $p_1^2=m_s^2$ and $p_2^2=p_3^4=p_4^2=0$,
\begin{equation}
\begin{split}
&\int \frac{{\rm d }^3 p_4 }{2\,E_4} F(p_1,p_2,p_3,p_4)\delta(p_1+p_2-p_3-p_4)=\\
&\int {\rm d }^4 p_4 \, \delta(p_4^2)\delta(p_1+p_2-p_3-p_4) F(p_1,p_2,p_3,p_4)=\\
&F(p_1,p_2,p_3,p_1+p_2-p_3)\delta(m_s^2+2p_1\cdot p_2-2p_1\cdot p_3-2p_2\cdot p_3)
\end{split}
\end{equation}

Also, we used some notable integrals:
\begin{itemize}
\item[i)
]\begin{equation}
\int{\rm d }x \frac{1}{\left[a+b\,x\right]^2}=-\frac{1}{b(a+bx)}+const.\,,
\label{int1}
\end{equation}
implying that
\begin{equation}
\int_{-1}^{+1}{\rm d }x \frac{1}{\left[a+b\,x\right]^2}=\frac{2}{a^2-b^2}\,.
\label{int1b}
\end{equation}
\item[ii)]
\begin{equation}
\int{\rm d }x \frac{(1+x)^2}{\left[a+b\,x\right]^4}=-\frac{a^2+a(3b\,x+b)+b^2(3x^2+3x+1)}{3b^3(a+bx)^3}+const.\,,
\label{int2}
\end{equation}
implying that
\begin{equation}
\int_{-1}^{+1}{\rm d }x \frac{(1+x)^2}{\left[a+b\,x\right]^4}=\frac{8}{3(a-b)(a+b)^3}=\frac{8}{3}\frac{1}{(a^2-b^2)(a+b)^2}\,.
\label{int2b}
\end{equation}
\item[iii)] 
\begin{equation}
\int{\rm d }x \frac{(1+x)}{\left[a+b\,x\right]^3}=-\frac{a+2b\,x+b}{2b^2(a+bx)^2}+const.\,,
\label{int3}
\end{equation}
implying that
\begin{equation}
\int_{-1}^{+1}{\rm d }x  \frac{(1+x)}{\left[a+b\,x\right]^3}=\frac{2}{(a-b)(a+b)^2}\,.
\label{int3b}
\end{equation}
\end{itemize}
%%%%%%%%%%%%%
\section{Numerical reduction of the decay and collisional integrals}\label{appB}
%%%%%%%%%%%%%
In this Appendix we show how the integrals in Eq.~(\ref{general_coll_integral}) can be reduced from nine to three dimensions (four in the case of the decay processes) using the procedure reported in Ref.~\cite{Hannestad:1995rs}. Although the procedure is not new, we recall it here for completeness.\\
Using the property
\begin{equation}
\frac{{\rm d}^3p_4}{2E_4}={\rm d}^4p_4\delta (p_4^2-m_4^2)\Theta(p_4^0)\,\ ,\label{deltaprop}
\end{equation} 
the integral over $p_4$ is done using the delta in Eq.~(\ref{general_coll_integral}).  For the scattering processes, we obtain:
\begin{equation}
p_4=p_1+p_2-p_3 \,\ .
\label{p4}
\end{equation}
Introducing the following angles
\begin{align}
\cos (\alpha)&=\frac{\mathbf{p_1\cdot p_2}}{p_1p_2} \,\ ,\\
\cos (\theta)&=\frac{\mathbf{p_1\cdot p_3}}{p_1p_3} \,\ ,\\
\cos (\alpha ')&=\frac{\mathbf{p_2\cdot p_3}}{p_2p_3}=\cos\alpha\cos\theta+\sin\alpha\sin\theta\cos\beta\,\ ,
\end{align}
we can write
\begin{align}
{\rm d}^3p_2&=p_2^2{\rm d}p_2{\rm d}\cos\alpha{\rm d}\beta \,\ ,\\
{\rm d}^3p_3&=p_3^2{\rm d}p_3{\rm d}\cos\theta {\rm d}\mu \,\ , 
\end{align}
with $\beta$ and $\mu$ the azimuthal angles for $\mathbf{p_2}$ and $\mathbf{p_3}$.
The integration over $d\beta$ is carried out using the $\delta$ function:
\begin{equation}
p_4^2-m_4^2=f(\beta) \,\ .
\end{equation}
We use the relation for the $\delta$:
\begin{equation}
\int {\rm d}\beta\delta(f(\beta))=\sum_i\int {\rm d}\beta \frac{1}{\big|{\rm d}f(\beta)/{\rm d}\beta\big|_{\beta=\beta_i}}\delta(\beta-\beta_i) \,\ ,
\end{equation}
where the $\beta_i$ are the roots of $f(\beta)=0$. Using the previously introduced angles
\begin{equation}
\frac{{\rm d}f(\beta)}{{\rm d}\beta}=2p_2p_3\sin\alpha\sin\theta\sin\beta \,\ ,
\end{equation}
$\sin\beta_i$ is found as $\pm(1-\cos^2\beta_i)^{1/2}$, where
\begin{equation}
\cos\beta_i=\frac{2E_2E_3-2p_2p_3\cos\alpha\cos\theta-Q-2E_1E_2+2p_1p_2\cos\alpha+2E_1E_2-2p_1p_3\cos\theta}{2p_2p_3\sin\alpha\sin\theta} \,\ ,
\end{equation}
and $Q\equiv m_1^2+m_2^2+m_3^2-m_4^2$. The equation for $\cos\beta$ has two solutions, but we can account for them by multiplying by two and using as integration's interval $[0,\pi]$.
The limits of integration in ${\rm d}\cos\alpha$ come from demanding that $\cos^2\beta\leq 1$, meaning that
\begin{equation}
(2p_2p_3\sin\alpha\sin\theta\sin\beta)^2\geq 0 \,\ .
\end{equation}
This is the same requirement that $({{\rm d}f(\beta)}/{{\rm d}\beta})^2\geq 0$. Therefore we can write
\begin{equation}
\int_0^{2\pi}{\rm d}\beta\delta(f(\beta))=2\frac{1}{|{\rm d}f(\beta)/{\rm d}\beta|_{\beta=\beta_i}^2}\Theta\Bigg(\Bigg|\frac{{\rm d}f(\beta)}{{\rm d}\beta}\Bigg|_{\beta=\beta_i}^2\Bigg) \,\ .
\end{equation}
Introducing the following definitions:
\begin{align*}
\gamma&=E_1E_2-E_1E_3-E_2E_3 \,\ ; \\
\epsilon&=p_1p_3\cos\theta \,\ ; \\
k&=p_1^2+p_3^2 \,\ ; \\
a&=p_2^2(-4k+8\epsilon) \,\ ; \\
b&=p_2(p_1-\epsilon/p_1)(8\gamma+4Q+8\epsilon) \,\ ; \\
c&=-4\gamma^2-4\gamma Q-Q^2-8\gamma\epsilon-4Q\epsilon-4\epsilon^2+4p_2^2p_3^2(1-\cos\theta)^2 \,\ ;
\end{align*}
 the derivative can be written as:
\begin{equation}
\bigg|\frac{{\rm d}f(\beta)}{{\rm d}\beta}\bigg|_{\beta=\beta_i}=\sqrt{a\cos^2\alpha+b\cos\alpha+c} \,\ .
\end{equation}
All possible matrix elements only include products of the four-momenta. All the products are analytically integrable over ${\rm d}\cos\alpha$ and can be carried out by using these relations:
\begin{align*}
\int\frac{1}{\sqrt{ax^2+bx+c}}\Theta(ax^2+bx+c){\rm d}x&=\frac{\pi}{\sqrt{-a}}\Theta(b^2-4ac)\,\ ; \\
\int\frac{x}{\sqrt{ax^2+bx+c}}\Theta(ax^2+bx+c){\rm d}x&=-\frac{b}{2a}\frac{\pi}{\sqrt{-a}}\Theta(b^2-4ac)\,\ ; \\
\int\frac{x^2}{\sqrt{ax^2+bx+c}}\Theta(ax^2+bx+c){\rm d}x&=\Bigg(\frac{3b^2}{8a^2}-\frac{c}{2a}\Bigg)\frac{\pi}{\sqrt{-a}}\Theta(b^2-4ac)\,\ . 
\end{align*} 
The step function comes from demanding a real integration interval. This also ensures that the roots of $ax^2+bx+c$ are not outside the fundamental integration interval of $[-1,1]$. Integration over ${\rm d}\mu$ is trivial because there is no dependence on this parameter.\\
All the possible products of these momenta are calculated below:
\begin{align*}
p_1\cdot p_2&=E_1E_2-p_1p_2\cos\alpha \,\ , \\
p_1\cdot p_3&=E_1E_3-p_1p_3\cos\theta \,\ , \\
p_1\cdot p_4&=m_1^2+(E_1E_2-p_1p_2\cos\alpha)-(E_1E_3-p_2p_3\cos\theta) \,\ , \\
p_2\cdot p_3&= (E_1E_2-p_1p_2\cos\alpha)-(E_1E_3-p_1p_3\cos\theta)+\frac{Q}{2} \,\ , \\
p_2\cdot p_4&=(E_1E_3-p_1p_3\cos\theta)+m_2^2-\frac{Q}{2} \,\, \\
p_3\cdot p_4&=(E_1E_2-p_1p_2\cos\alpha)-m_3^2+\frac{Q}{2} \,\ . 
\end{align*}
To integrate over ${\rm d}\cos\theta$, the solutions of $b^2-4ac$ are important for the the integration interval:
\begin{equation}
\cos\theta=\frac{-2\gamma-2p_2^2-Q\pm2p_2\sqrt{2\gamma+p_1^2+p_2^2+p_3^2+Q}}{2p_1p_2} \,\ .
\end{equation}
If there is to be a real integration interval, both of these solutions must be real and we will refer to them as $\cos\theta_{\mathrm{min}}$ and $\cos\theta_{\mathrm{max}}$. The real integration limits are $\alpha=\sup[-1,\cos\theta_{\rm min}]$ and $\beta=\inf[+1,\cos\theta_{\rm max}]$ with $\alpha\leq\beta$.
Finally, with these conditions, it is possible to calculate numerically the collision integral left:
\begin{equation}
C_{\rm coll}[f]=\frac{2}{(2\pi)^4}\frac{1}{2E_1}\int_0^{\infty}\int_0^{p_1+p_2}\int_{\alpha}^{\beta}\frac{p_2^2{\rm d}p_2}{2E_2}\frac{p_3^2{\rm d}p_3}{2E_3}{\rm d}\cos\theta F(f_i,f_f)\Lambda(p_1,p_2,p_3)\Theta(A) \,\ ,
 \end{equation} 
where $A$ is the parameter space allowed, $\Lambda$ comes from the following analytical integral:
\begin{equation}
\Lambda(p_1,p_2,p_3)\equiv \int\frac{|M|^2}{\sqrt{a\cos^2\alpha+b\cos\alpha+c}}\Theta(a\cos^2\alpha+b\cos\alpha+c){\rm d}\cos\alpha \,\ ,
\end{equation}
and $F(f_i,f_f)$ is the expression defined in Eq.~(\ref{Ffactor}) and evaluated  for $p_4$ given in Eq.~(\ref{p4}).

Finally, for the decay processes, it is possible to follow a similar procedure modulo the fact that we do not integrate analytically over $\cos\alpha$ and we define:
\begin{align}
p_4&=p_1-p_2-p_3 \,\ ,\\
\gamma_{\rm dec}&\equiv E_2E_3-E_2E_1-E_3E_1 \,\ , \\
b_{\rm dec}&\equiv 4p_2\left(-p_1+\epsilon/p_1\right)\left(2\gamma+Q+2e\right) \,\ .
\end{align}
%%%%%%
\section{Comparison with previous results}
\label{appC}
\begin{figure}[h]
\centering
    \includegraphics[scale=0.6]{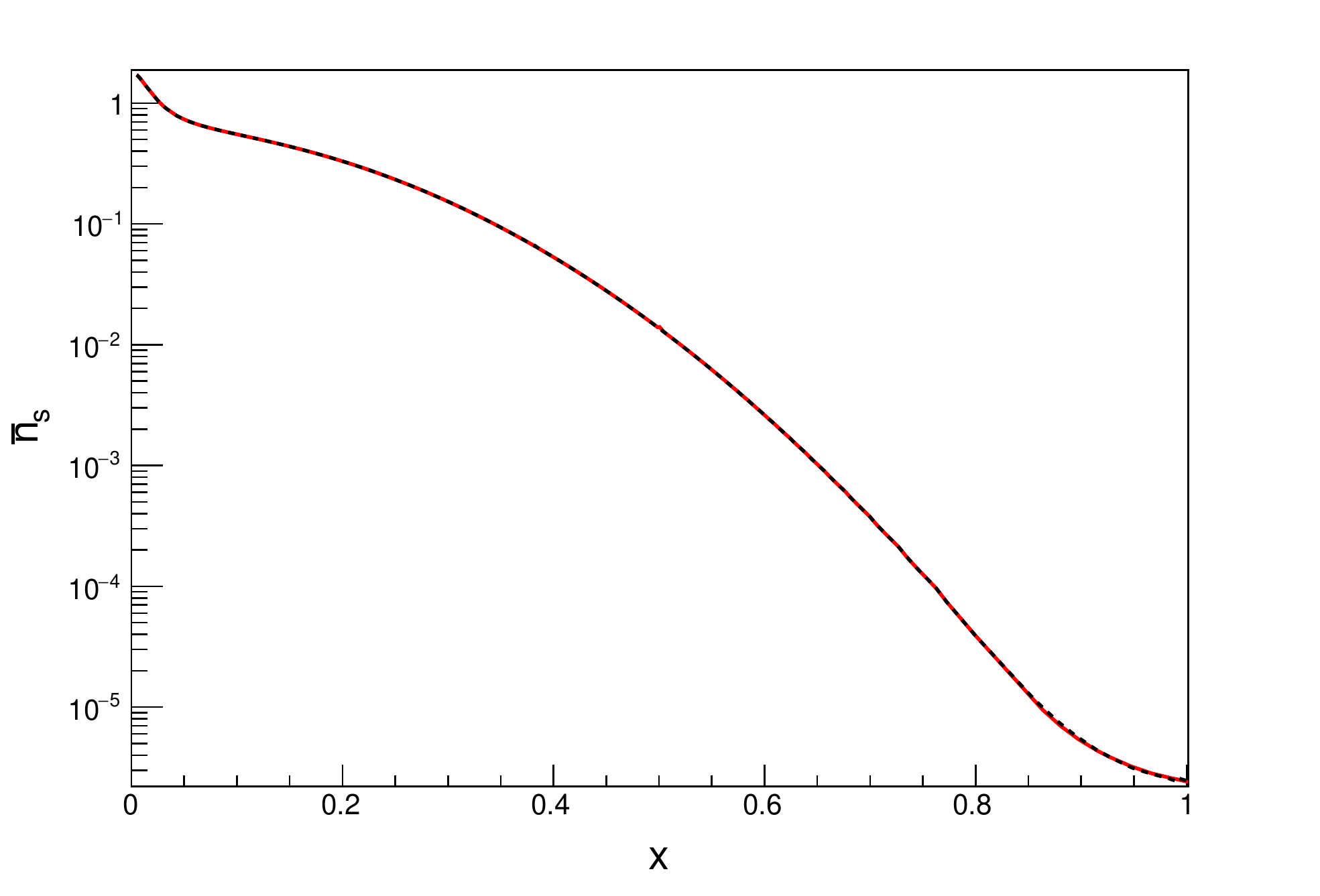}
    \caption{Comparison between our results (red, solid) and Ref.~\cite{Dolgov:2000jw} (black, dashed) on the comoving density of the sterile species for $m_s=100~\mathrm{MeV}$ and $\tau_s=0.055~\mathrm{s}$.}
    \label{ConfrontoDolgov}
\end{figure}

%%%%%%%
To validate our code, we have first compared our results with pioneering results obtained in ref.~\cite{Dolgov:2000jw}, finding excellent agreement with respect to several outputs, see e.g. Fig~\ref{ConfrontoDolgov} showing the comoving density of the sterile species.   

%%%%%%%%%%%%%%%%%%%%%%%%%%%%%%%%%%%%%%%%%%%%%%%%%%%%%%%%%%%%%%%%%%%%%%%%%%%%%%%%%%%%%%
%%%%%%%%%%%%%%%%%%%%%%%%%%%%%%%%%%%%%%%%%%%%%%%%%%%%%%%%%%%%%%%%%%%%%%%%%%%%%%%%%%%%%%%

%%%%%%%%%%%%%%%%%%%%%%%%%%%%%%%%%%%%%%%%%%%%%%%%%%%%%%%%%%%%%%%%%%%%%%
\begin{figure}
\centering
\includegraphics[scale=0.6]{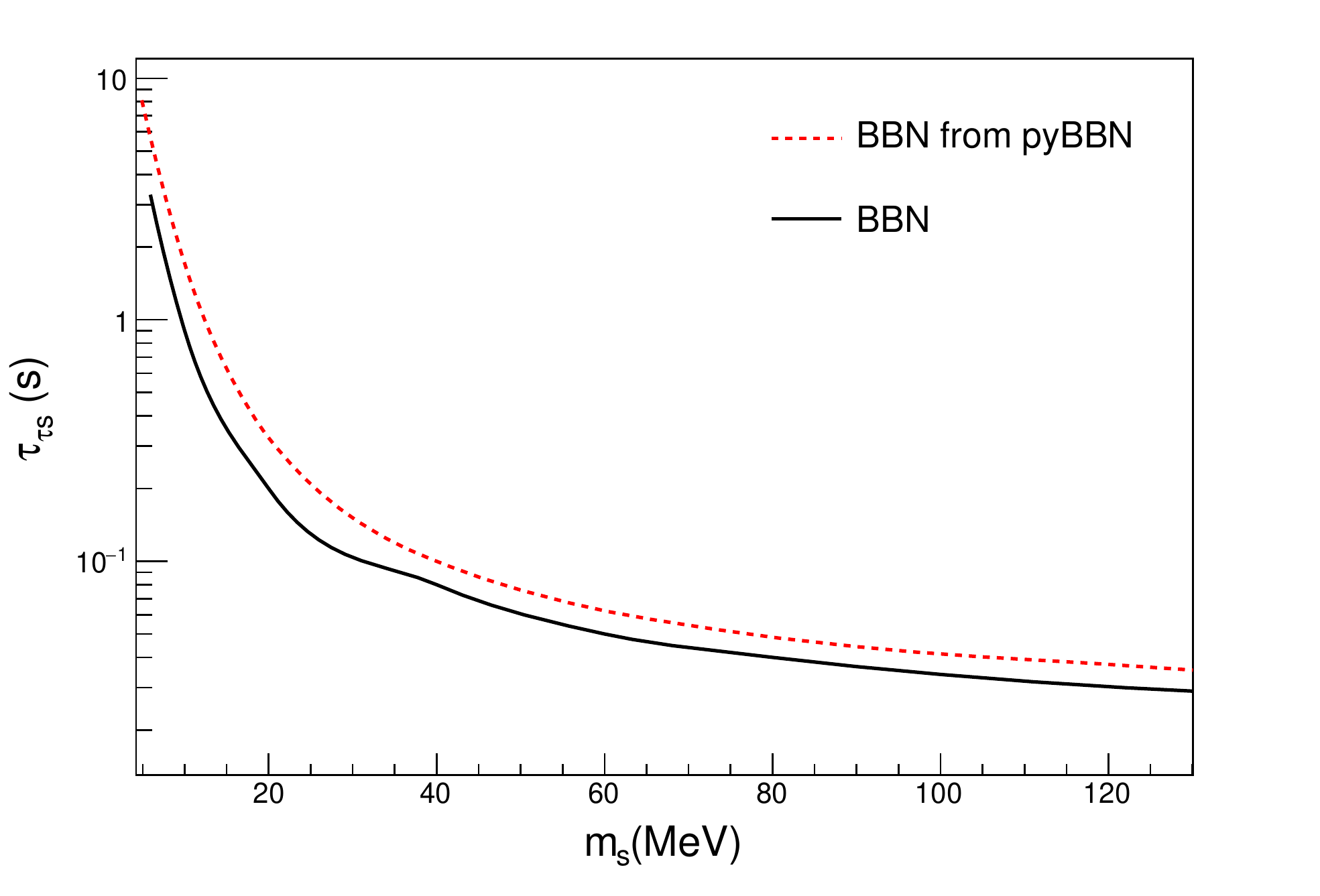}
\caption{Comparison of results from the BBN constraints for the decay time of a sterile neutrino mixed only with active tauonic (or muonic) neutrino in Ref.~\cite{Sabti:2020yrt,Dolgov:2000jw} and the one from our code.}
\label{comparison_BBN_t}
\end{figure}
%%%%%%%%%%%%%%%%%%%%%%%%%%%%%%%%%%%%%%%%%%%%%%%%%%%%%%%%%%%%%%%%%%%%%%
\begin{figure}
\centering
\includegraphics[scale=0.6]{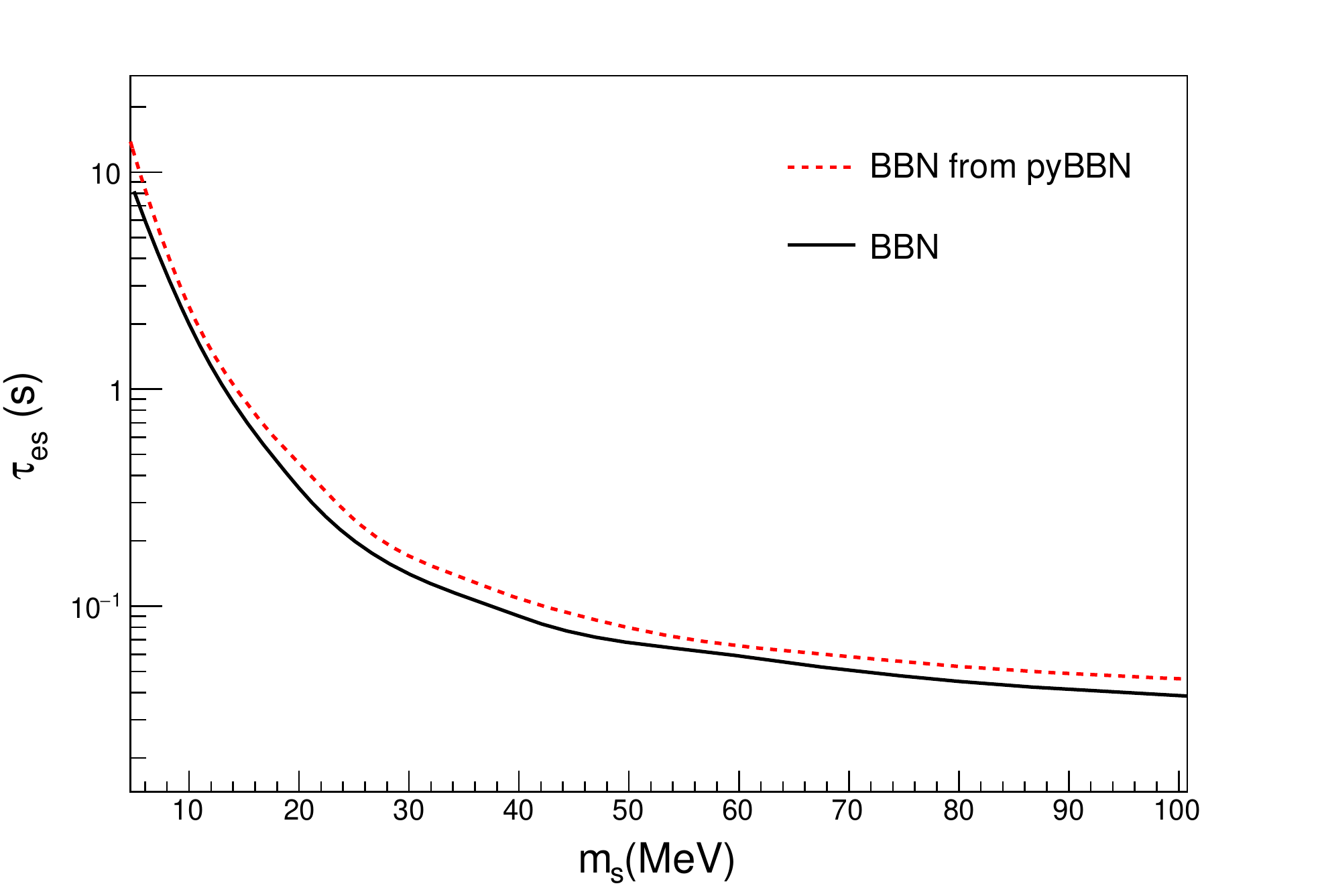}
\caption{Comparison of results from the BBN constraints for the decay time of a sterile neutrino mixed only with active electron neutrino in Ref.~\cite{Sabti:2020yrt,Dolgov:2000jw} and the one from our code.}
\label{comparison_BBN_e}
\end{figure}

%%%%%%%%%%%%%%%%%%%%%%%%%%%%%%%%%%%%%%%%%%%%%%%%%%%%%%%%%%%%%%%%%%%%%%
\begin{figure}
\centering
\includegraphics[scale=0.6]{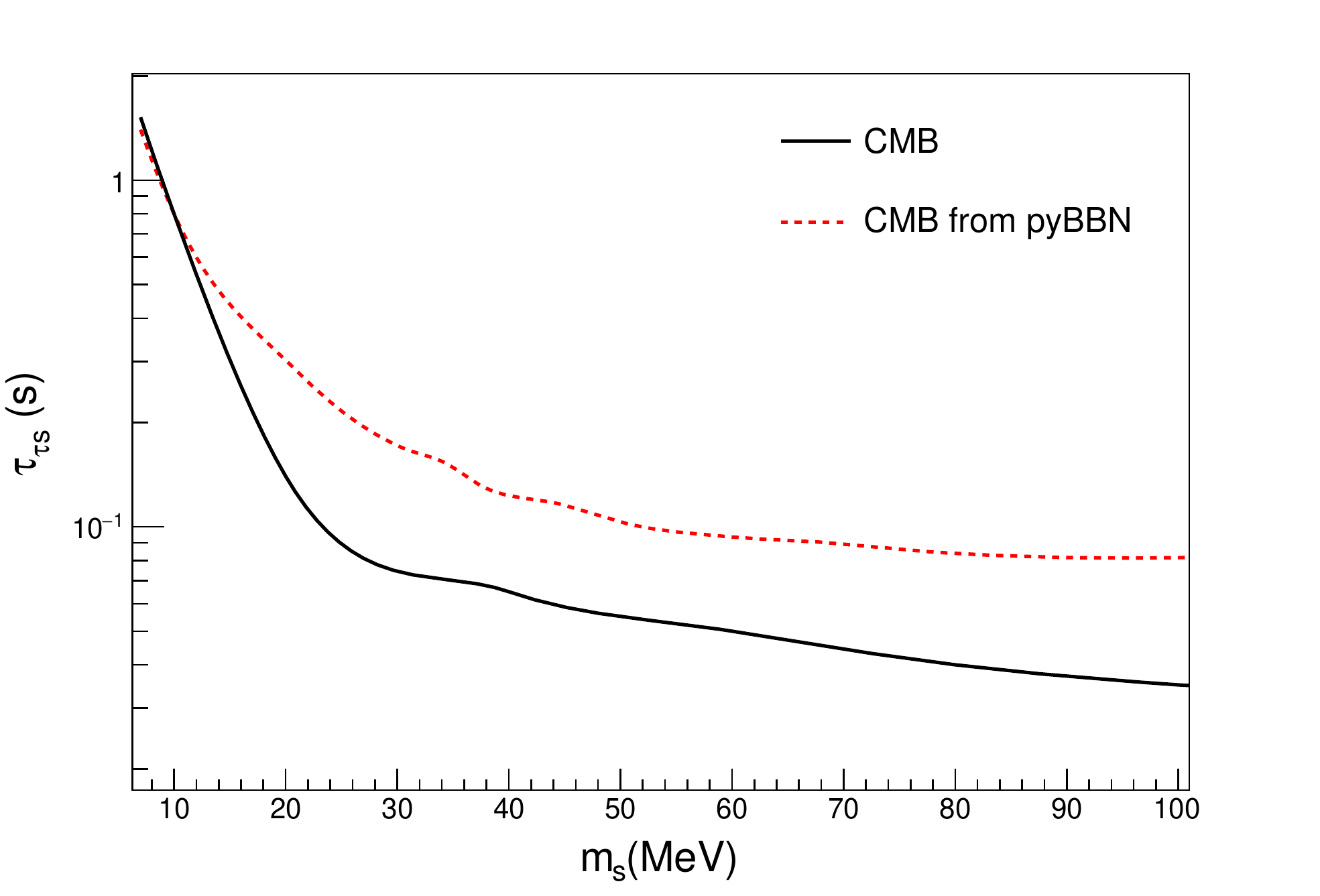}
\caption{Comparison of results from the CMB constraints for the decay time of a sterile neutrino mixed only with active tauonic (or muonic) neutrino in Ref.~\cite{Sabti:2020yrt} and the one from our code obtained using their same value of $\Theta_{\mathrm{Obs}}$ as a benchmark.}
\label{comparison_CMB_t}
\end{figure}
%%%%%%%%%%%%%%%%%%%%%%%%%%%%%%%%%%%%%%%%%%%%%%%%%%%%%%%%%%%%%%%%%%%%%%
\begin{figure}
\centering
\includegraphics[scale=0.6]{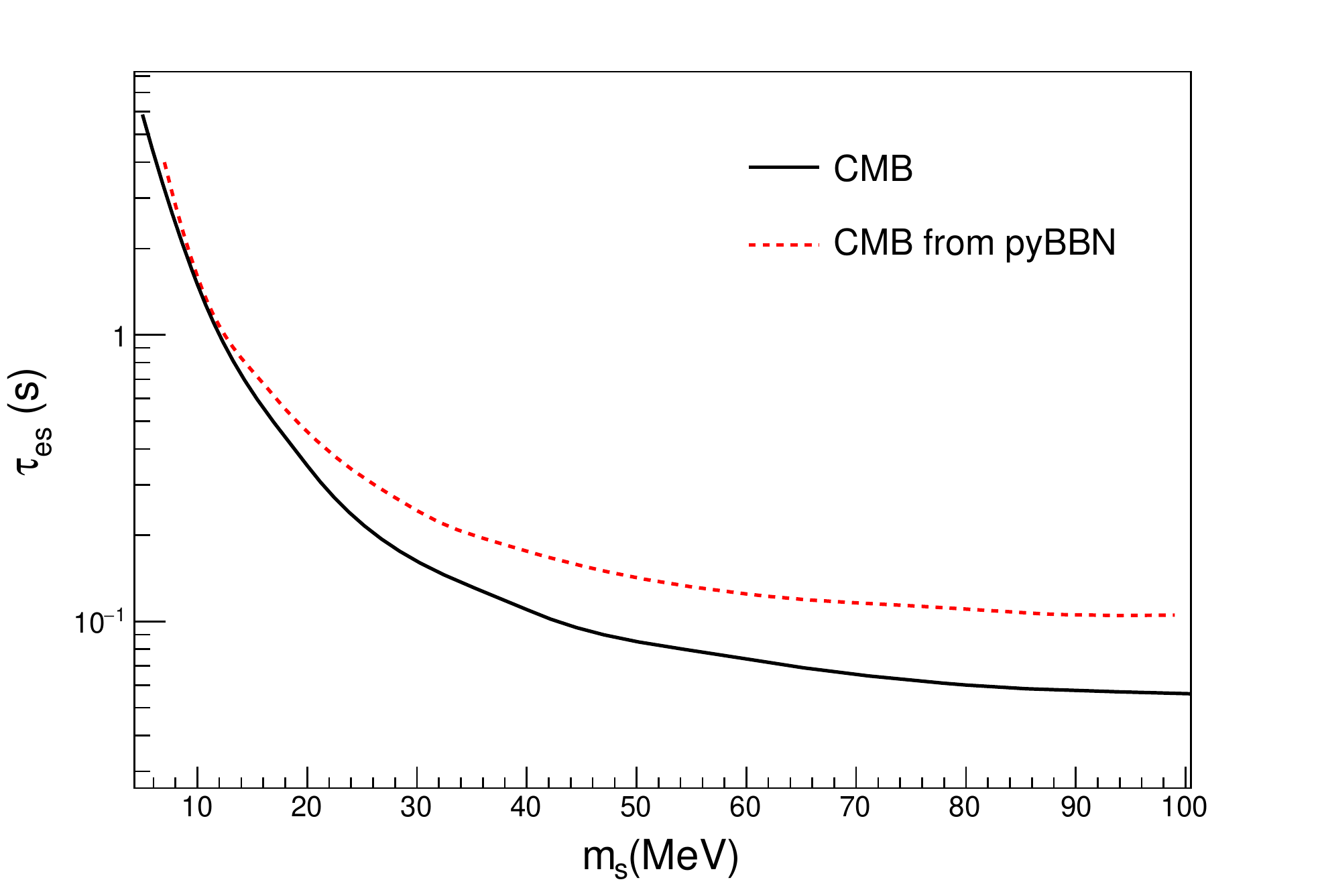}
\caption{Comparison of results from the CMB constraints for the decay time of a sterile neutrino mixed only with active electron neutrino in Ref.~\cite{Sabti:2020yrt} and the one from our code obtained using their same value of $\Theta_{\mathrm{Obs}}$ as a benchmark.}
\label{comparison_CMB_e}
\end{figure}

The most recent constraints on heavy decaying sterile neutrinos have been reported in Ref.~\cite{Sabti:2020yrt}. 
In Fig.~\ref{comparison_BBN_t} and \ref{comparison_BBN_e} we compare BBN bounds from Ref.~\cite{Sabti:2020yrt} with our BBN bounds, while in 
 Fig.s~\ref{comparison_CMB_t} and \ref{comparison_CMB_e} we compare CMB bounds from Fig.~11 in Ref.~\cite{Sabti:2020yrt} with CMB results from our code, obtained using the same value of $\Theta_{\mathrm{Obs}}$ as a benchmark.
 
While there is always a qualitative agreement, the quantitative agreement between the results is rather good only for the BBN case, while
it shows some discrepancy in the CMB case for high masses. From inspection of ref.~\cite{Sabti:2020yrt} (e.g., Section 4.1.2, Appendix B) we infer that the authors
find a systematically lower value of $N_{\rm eff}$ than us when masses are significantly large than $\sim 10\,$MeV, and that their $N_{\mathrm{eff}}$ can  also attain values below 3, while we always find $\Delta N_{\rm eff}\gtrsim 0$ (see e.g. Fig.~\ref{Neff_100vs30}). This has been confirmed by private communication with the corresponding author of ref.~\cite{Sabti:2020yrt}. 
 This behaviour is also found in Ref.~\cite{Ruchayskiy:2012si} (right panel of figure 3) and is indeed traced back by the authors of~\cite{Sabti:2020yrt} to the solution scheme provided by the \texttt{pyBBN} code. As a result, in~\cite{Sabti:2020yrt} CMB bounds at large masses are dominated by $Y_p$ (to which CMB is less sensitive) rather than by $N_{\mathrm{eff}}$, and are weaker than ours.
Since BBN bounds from $Y_p$ depend mostly on spectral distortions of the electron-type neutrinos, it is not surprising that the agreement is much better in this observable channel. For the case of mixing with $\nu_\tau$, where $N_{\rm eff}$ plays a slightly bigger role, the agreement is not as excellent while remaining good.  Note that the inclusion of deuterium constraints is subleading to the effect on $Y_p$ (see Fig. 3 in~\cite{Sabti:2020yrt}), so neglecting it has no major impact on our BBN bounds.

In the recent paper~\cite{Boyarsky:2021yoh}, a physical interpretation of the effect on $N_{\rm eff}$ found in~\cite{Sabti:2020yrt} is discussed: It is claimed that  $\Delta N_{\rm eff}<0$ is a quasi-generic outcome of injection of energy after neutrino decoupling, even for decay modes mostly in neutrinos, as a result of a dominant entropy transfer to $e^\pm$ via non-thermal neutrino interactions with particles of the thermal bath, see Eqs.~3, 4, 5 in~\cite{Boyarsky:2021yoh}.  Although the authors of~\cite{Boyarsky:2021yoh} draw some analogy of their   $\Delta N_{\rm eff}<0$ effect with the results of the pioneering work~\cite{Hannestad:2004px} on a low-reheating scenario, we believe that this is not very instructive, since the thermal neutrino bath is obviously suppressed if the universe starts at temperatures comparable or lower than the neutrino decoupling, a situation very different from the one under study. 

Since we could not confirm numerically these surprising results, we thought useful to discuss their plausibility with a qualitative study of the Boltzmann equation, in the analytical approximation derived above, matching e.g. the one in~\cite{Dolgov:2000jw}. If denoting with $f_1$ the non-thermal neutrino distribution under exam, it obeys an equation of the form: 
\begin{equation}
x\partial_x f_1=\frac{I[f_1]}{H}=\frac{1}{H}\left[S(x)+\varsigma^2 G_F^{2}\left(f^{\mathrm{eq}}-f_1\right)T^4E_1\right]\,.\label{appexpl}
\end{equation}
The first term $S\propto (f_s-f_s^{eq})/\tau_s\simeq f_s/\tau_s$ at the r.h.s is always positive, since describing the injection due to decays of the sterile neutrinos. More precisely, $S$ is an integral where $(f_s-f_s^{eq})$ enters as a kernel, see e.g.~Eq.~(\ref{apprsouterm}).
The second (collisional) term, where $\varsigma^2$ is a positive numerical constant,  is initially zero since neutrinos are at equilibrium, but as a result of the source term $S$, $f_1$ grows above the equilibrium value and leads to a negative value of the collisional term, linear in $(f^{\mathrm{eq}}-f_1)$. 
A depletion of the neutrino distribution to ``sub-thermal'' values, as apparently found in~\cite{Boyarsky:2021yoh}, would imply reversing the sign of the collisional term, i.e. obtaining  a {\it positive} value of $(f^{\mathrm{eq}}-f_1)$. However, this second term being controlled by the source term $S$, it can at most grow negative to the point of compensating the first term, thus reaching an equilibrium between injection and collisional redistribution of the energy. At that moment, the derivative of $f_1$ is driven to zero and $f_1$ becomes constant.
In practice, unless none of the rates is fast compared to the Hubble expansion $H$, the evolution follows one of the following paths:

\begin{itemize}
\item[i)] The first term at the r.h.s of Eq.~(\ref{appexpl}) dominates over the collisional term, which means that $f_1$ grows significantly larger than  $f^{\mathrm{eq}}$, with the scattering incapable of fully compensating it.

\item[ii)] The second term at the r.h.s of Eq.~(\ref{appexpl}) is dominant:  As a result,  $f_1$ tends to $f^{\mathrm{eq}}$, annihilating the collisional term, or more precisely settling to a slightly larger-than-thermal value to compensate for the injection. 
\end{itemize}
Hence, we conclude that either $f_1$ freezes out at a value larger than the equilibrium one (implying  $\Delta N_{\rm eff}>0$), or in the `worst' case it tends to the equilibrium distribution ( $\Delta N_{\rm eff}\to 0^+$), in contradiction with the conclusions of ref.~\cite{Boyarsky:2021yoh}. 
In the above discussion, we neglected the effect of the growing temperature as a result of the transfer of entropy from the sterile neutrino decays: At this level of approximation, however, its effect is to make the collisional term increase via the $T^4$ factor, as well as to increase $f^{\mathrm{eq}}$, so that the equilibrium distribution the particles are driven to is not the same as the initial one.

\begin{figure}[h]
\centering
    \includegraphics[scale=0.6]{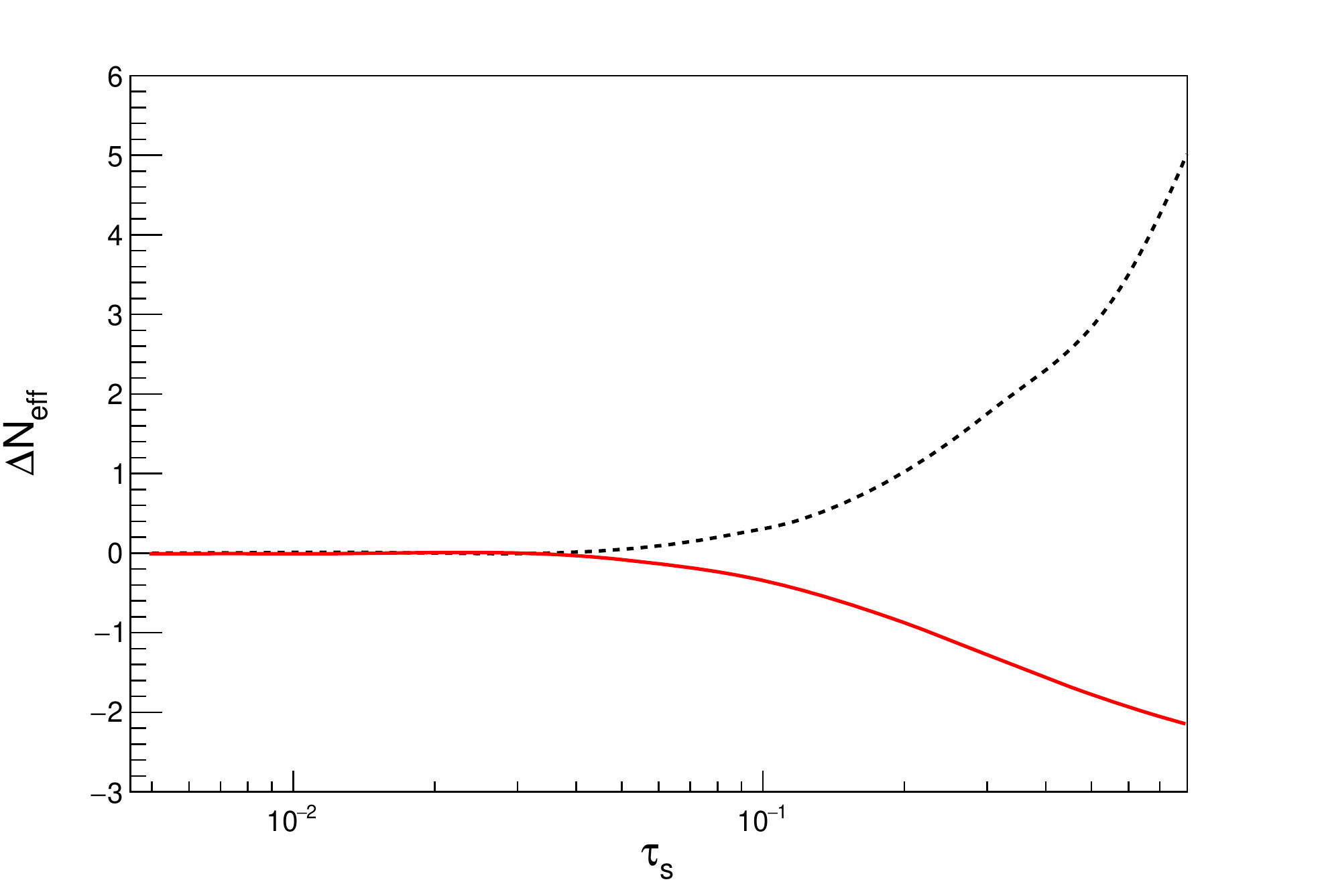}
    \caption{$\Delta N_{\rm eff}$ vs. $\tau_s$ for 
    $m_s=50~\mathrm{MeV}$ in the limit of zero b.r. into active neutrino states (solid red line) and zero b.r. into electromagnetic sector (black dashed line).}
    \label{negDNeff}
\end{figure}

For the sake of the argument, let us assume that as a result of collisions, the second term at r.h.s of Eq.~(\ref{appexpl})  starts becoming positive at some instant, i.e. $f^{\mathrm{eq}}>f_1$: Then, $f_1$ would start growing again (because the r.h.s would be positive) and hence the r.h.s. is brought closer to zero. The feedback is such that any collisionally induced depletion of neutrinos would be immediately compensated, preventing a depletion of $f_1$ in the circumstances under exam. For that to happen, one needs a situation in which  $f^{\rm eq}$ grows due to the growth of temperature, with a minimal impact on $f_1$. This can be easily obtained if the injection of energy
from decay is dominantly in the e.m. sector and the collisional terms is negligible, i.e. a situation of type  i) but in the e.m. sector, as naively expected. We checked that this is indeed the case, namely one obtains $\Delta N_{\rm eff}<0$ when artificially pushing the b.r. into neutrinos to sub-dominant values, and the decay happens sufficiently late. These regimes are illustrated in Fig.~\ref{negDNeff}:
If the sterile state decays early, the collisional term is capable of restoring equilibrium and $\Delta N_{\rm eff}\to 0$, as argued in the situation ii) described above, independently of the channel in which energy is injected. For a  decay happening later and later, collisions are less and less efficient in restoring equilibrium and $\Delta N_{\rm eff}> 0$ for a dominant b.r. into neutrino states, while $\Delta N_{\rm eff}< 0$ for a dominant b.r. into the e.m. channel.

%%%%%%%%%%%%%%%%%%%%%%%%%%%%%%%%%%%%%%%%%%%%%%%%%%%%%%%%%%%%%%%%%%%%%%%%%%%%%%%%%%%%%%
%%%%%%%%%%%%%%%%%%%%%%%%%%%%%%%%%%%%%%%%%%%%%%%%%%%%%%%%%%%%%%%%%%%%%%%%%%%%%%%%%%%%%%%

All our numerical results qualitative agree with these conclusions that can also drawn from Eq.~(\ref{appexpl}). Since Eq.~(\ref{appexpl}) is an approximation, one may wonder if the results of ref.~\cite{Boyarsky:2021yoh} are due to some features not captured by Eq.~(\ref{appexpl}) (and, for some unknown reason, also missed by our numerical calculation). We believe that this is not the case as we argue in the following:

\begin{itemize}
\item Eq.~(\ref{appexpl}) neglects quantum statistics effects, which appear however irrelevant to the arguments of ref.~\cite{Boyarsky:2021yoh}, and are anyway present for both electrons and neutrinos, without causing a qualitative change in one sector compared to the other.

\item Eq.~(\ref{appexpl}) assumes $T_\nu=T$. However, this is not crucial to the conclusions above. The presence of a ``two temperatures background'' is essentially equivalent to split the term $(f^{\mathrm{eq}}-f_1)$ into a linear combination of $(f_\nu^{\mathrm{eq}}-f_1)$ and  $(f_e^{\mathrm{eq}}-f_1)$, each weighted by a positive factor. Both these functions are negative under the effect of neutrino energy injection:  One has {\it a fortiori} $(f_\nu^{\mathrm{eq}}-f_1)< 0$ since  $f_\nu^{\mathrm{eq}}< f_e^{\mathrm{eq}}$  when $T>T_\nu$ (and electrons are relativistic), while $(f_e^{\mathrm{eq}}-f_1)< 0$ must be satisfied if we ask for $T$ to grow larger than $T_\nu$ as a result of injection of energy in the neutrino sector. The $T$ evolution equation is indeed controlled by the {\it opposite} of the neutrino collisional term, i.e. goes as $(f_1-f_e^{\mathrm{eq}})$, as can be checked via Eq.~(\ref{temperature evolution}).

\item Eq.~(\ref{appexpl}) {\it does not} account for the quadratic terms involving the non-thermal parts of the neutrino distributions, i.e. depending from $(f_\nu-f_\nu^{\rm eq})^2$, since thermal distributions have been used in the kernels of the collisional integrals, see Appendix~\ref{appA}. Note however that these {\it are not} the processes claimed in~\cite{Boyarsky:2021yoh} to be responsible for the effect on $N_{\rm eff}$, since their Eq.s 3, 4, 5 explicitly indicate reactions of the non-thermal neutrinos on thermal background species, which {\it are included} in the above treatment. Yet, let us entertain the possibility that the results outlined by the authors of~\cite{Boyarsky:2021yoh} are physical and due to these non-linear effects, and that it is only their interpretation/attribution to be incorrect. One should then expect that  $\Delta N_{\rm eff}<0$ shows a
 ``threshold'' behaviour with respect to the amount of energy injected in neutrinos: The less non-thermal energy is injected, the less likely these non-linear interactions among non-thermal particles should become when compared to interactions with the thermal background. Hence, the authors of~\cite{Boyarsky:2021yoh} should have found that the effect kicks in only above some fraction of the background energy injected in the medium, growing quadratically above this value. Instead, not only they claim an effect also in the idealized case of `single neutrino injection' (see their Fig. 1), but clearly show an effect that is roughly linear in the injected non-thermal energy (see Fig. 7, left: The change is from slightly below -0.2\% to about -0.8\% when moving from an injection of 1\% to 5\%), inconsistent with this hypothetical explanation.
\end{itemize}

In conclusion, to the best of our knowledge, we attribute our departure from the results in~\cite{Sabti:2020yrt} on the CMB constraints at large masses to some unidentified systematics, probably the same effects responsible for the ``universal'' $\Delta N_{\rm eff}<0$ outcome described in~\cite{Boyarsky:2021yoh} which appears unphysical, for the reasons detailed above. The toy model advocated in~\cite{Boyarsky:2021yoh} to support their findings is also untrustworthy, since it does not account for reverse reactions (only reactions transferring energy from neutrinos into e.m. particles are included) and does not implement the physical requirement that only excess energy (above the thermal value) can be effectively transferred in collisions. As a consequence, it is for instance incapable of predicting equilibration and $\Delta N_{\rm eff}\to 0$ when energy is injected at early times.

{\it Note added:} After our paper was accepted for publication, a new version of ref.~\cite{Boyarsky:2021yoh} appeared including an appendix D where the authors criticise the generality of the conclusions reported above.  We believe that their arguments are incomplete and incorrect, as explained below.

First, they claim that our discussion based on Eq.~(\ref{appexpl}) and leading to the conclusion that $\Delta N_{\rm eff}\geq 0$ only applies to a situation close to equilibrium, i.e. at high-$T$. The authors seem to believe that we only consider a limiting condition i) where the solution to Eq.~(\ref{appexpl}) writes
\begin{equation}
f(E, x)=f^{\mathrm{eq}}+\frac{S(E, x)}{\varsigma^2 G_F^{2}T^4E}\,,
\end{equation}
which holds when {\it collisional processes are fast compared to the Hubble rate}, thus annihilating the r.h.s. of Eq.~(\ref{appexpl}). 

In fact, we also discuss the opposite limit ii) where 
{\it collisional processes are negligible compared to the Hubble rate}, i.e. when the solution after freeze-out $x_f$ can be expressed as
\begin{equation}
f(E, x)\simeq f^{\mathrm{eq}}+\int_{x_f}^{x}\frac{{\rm d}x}{xH}S(E, x)\,,
\end{equation}
 once again, leading to conclude that $f_\nu\geq f_{\rm eq}$ and hence $\Delta N_{\rm eff}\geq 0$.
iii) Finally, we present general arguments to prove why also for intermediate situations,  $\Delta N_{\rm eff}< 0$ is not to be expected. This is exactly based on the qualitative analysis of the collisional term describing the energy transfer between the neutrino and e.m. sector, claimed to be crucial in~\cite{Sabti:2020yrt} for situations of ``partial decoupling'' (i.e. around $T\sim 1\,$MeV).

To elucidate this point further, let us discuss what our analysis would conclude when applied to a concrete example. To ease the visualization of the contrasting conclusions, we choose to deal with the very schematic `compartments' system reported in the final paragraph of~\cite{Boyarsky:2021yoh}.
Consider the plasma as made of three populations: thermal neutrinos $f_\nu^\ell$, thermal electrons $f_e^\ell$, whose supports are essentially at low energies (hence the $\ell$ index) of order $T$, and the extra energetic neutrino flux $f_\nu^{h}$ coming from the decay. Without loss of generality, one may think of $f_\nu^{h}$ as a delta function at $E_{\rm inj}\gg T$. Of course, any modification of the initial thermal situation is driven by the latter component.
Note that, to draw conclusions on $N_{\rm eff}$, it is actually enough to refer to the integrated form of the Boltzmann equations referring to energy density which, in the relativistic limit, can be obtained by multiplying both members of equations like Eq.~(\ref{appexpl}) times $4\pi E^2$ and integrating over $E$. Since the temperature is a proxy for the energy density, this is not unlike what appears in Eq.~(\ref{temperature evolution}).  Finally, although we only report here Eq.~(\ref{appexpl}) for $f_\nu^{h}$, its weak scattering kernel is just the opposite of the sum of the weak scattering kernels of $f_e^\ell$ and $f_\nu^\ell$. Processes only redistributing the energy within each of the $\ell$-sectors disappear from the integrated form of the equations.

Since we are by hypothesis at moderate or low temperatures, the initial Ansatz is that the weak scattering among the $\ell$ particles is not very efficient and can be neglected; however, the weak scattering between the $h$ neutrinos and $\ell$ particles is still somewhat efficient, due to the large $E_{\rm inj}$, allowing for a (partial) redistribution of the injected energy. Till here, we share the analysis and conclusions of ref.~\cite{Boyarsky:2021yoh}. 
However, the finding of ref.~\cite{Boyarsky:2021yoh} that we question is that this redistribution is effective in transferring energy to the $\ell$-electron sector, but not so much in transferring energy to the $\ell$-neutrino sector. In practice, the $f_\nu^{h}$ would behave just as mediators that transfer most of their energy into the e.m. sector via weak scatterings, heating it without affecting the low-energy neutrinos.

Besides our numerical results, also an inspection of the Boltzmann equation drives us instead to very different conclusions.
 Just after the initial energy injection, clearly one has $f_e^\ell=f_\nu^\ell=f_{\rm eq}$: The weak scattering collisional terms for $f_\nu^\ell$ and $f_e^\ell$ depend on the same kernel, proportional to $T^4(f_\nu^{h}-f_{\rm eq})\sim u^\ell (f_\nu^{h}-f_{\rm eq})$, with $u^\ell$ being the energy density of the $\ell$-backgrounds. 
Even in the situation iii) where the $h-\ell$ scattering is partially effective, at early times the
injected energy is redistributed similarly to the neutrino and electron backgrounds, to the {\it pro rata} of their statistical and coupling factors (see Table~\ref{scattering}). If anything, a quantitative inspection reveals that the rescattering favours redistribution to the $\ell-$neutrino rather than to the $\ell$-electron sector. The only difference between $f_\nu^\ell$ and $f_e^\ell$ is that, while electrons rapidly redistribute their  gained energy via electromagnetic interactions to attain a new thermal distribution (this is why Eq.~(\ref{temperature evolution}) can be used, replacing the role of the Boltzmann equation for electrons), this process is only poorly effective among low-energy neutrinos. {\it But, integrated over energy, the overall transferred energies to the two species are comparable}, in contradiction with the findings of ref.~\cite{Boyarsky:2021yoh}.

At later times, once the $\ell$-distributions are distorted, no exact argument can be made but numerically. Nonetheless, as long as we neglect the feedback of the $\ell$-populations on the energy transfer from the $h$ one, the energy transferred to the electrons will depend on the energy integral of kernels roughly scaling as $u^\ell_e A\delta(E-E_{\rm inj})+ J(u^\ell_e f_\nu^\ell-u^\ell_\nu f_e^\ell)$, while the energy transferred to the neutrino sector will roughly depend on the kernel $\sim u_\nu^\ell B\delta(E-E_{\rm inj})+J( u_\nu^\ell f_e^\ell-  u_e^\ell f_\nu^\ell)$, with $B\gtrsim A >0$, and the terms proportional to $J(E)>0$ being negligible as long as one neglects the energy transfers among $\ell$-species.
These transfer terms are also vanishing at initial thermal conditions and, in an integral sense, stay small as long as the energies injected in the two sectors are comparable, which is the case at least at the beginning of non-thermal evolution, as argued above. More in general, while spectral distortions are possible and do take place, the role of these energy-exchange terms is to {\it counteract} distortions that would introduce an energy unbalance between the two sectors, not to generate or promote them, as ref.~\cite{Boyarsky:2021yoh} seems to imply.

When and how does the evolution stop? If the weak collisions are very efficient, there is eventually no more excess energy left in $f_\nu^h$, which corresponds to the limit where both neutrino and electron (or, more correctly, electromagnetic) background energies have been rescaled equally; then, $N_{\rm eff}\simeq 3$ follows, matching the expectation for a situation close to equilibrium despite the limitations of such a toy model: Not surprising,  since this is the meaning of a ``very efficient scattering regime''! In general, however, a residual extra energetic neutrino distortion $f_\nu^h$ will remain, albeit with a partially reprocessed energy spectrum. In this toy model, this would be the key responsible for a $\Delta N_{\rm eff}>0$ effect.  We thus recover qualitatively the same quantitative results illustrated in Fig.~\ref{negDNeff}, where the  $\Delta N_{\rm eff}$ is found to be a monotonic function of $\tau_s$ for the range of parameters in question.

\end{document}